\documentclass[12pt,preprint]{aastex}

\newcommand{\etal}{{et al.~}}

\shorttitle{Multiscale analysis in Orion A}
\shortauthors{Poidevin, F. \& Bastien, P.}

\begin{document}  %Do NOT delete this line

\title{Multi-scale analysis of magnetic fields in filamentary molecular clouds in Orion A}
\shortauthors{F. Poidevin \& P. Bastien}

% Type your authors as shown in the example below, listing each author in
% a separate author argument, with affiliation given in the affiliation argument
% as needed:

\author{Fr\'ed\'erick POIDEVIN}
\affil{Universidade de S\~{a}o Paulo, Instituto de Astronomia, Geof\'isica e C\^ien\c{c}as Atmosf\'ericas, 
Rua do Mat\~{a}o 1226, Butant\~{a}, S\~{a}o Paulo, SP 05508-900, Brazil}
\email{Poidevin@astro.iag.usp.br}

\author{P. Bastien}
\affil{D\'epartement de Physique and Observatoire du Mont-M\'egantic, Universit\'e de Montr\'eal, C.P. 6128, Succ. Centre-ville, Montr\'eal, Qu\'ebec H3C 3J7, Canada.}
\email{Bastien@astro.umontreal.ca}

\author{T. J. Jones \altaffilmark{1}}
\affil{University of Minnesota, Department of Astronomy, 116 Church St. S.E., 
Minneapolis, MN 55455}
\email{tjj@astro.umn.edu}

\altaffiltext{1}{Visiting Astronomer at the Infrared Telescope Facility which is operated 
by the University of Hawaii under contract from the National Aeronautics and Space Administration.}

\begin{abstract}
New visible and $K$-band polarization measurements on stars surrounding molecular clouds 
in Orion A and stars in the BN vicinity are presented. Our results confirm that magnetic fields 
located inside the Orion A molecular clouds and in their close neighborhood are spatially connected. 
On and around the BN object, we measured the angular offsets between the $K$-band polarization data 
and available submm data. We find high values of the polarization degree, $P_{K}$, and 
of the optical depth, $\tau_{K}$, close to an angular offset position of $90^{\circ}$ 
whereas lower values of $P_{K}$ and $\tau_{K}$ are observed for smaller angular offsets. 
We interpret these results as evidence for the presence of various magnetic 
field components toward lines of sight in the vicinity of BN.
On a larger scale, we measured the distribution of angular offsets between available $H$-band polarization data 
and the same submm data set. Here we find an increase of $<P_{H}>$ 
with angular offset which we interpret as a rotation of the magnetic field by $\lesssim 60^{\circ}$. 
This trend generalizes previous results on small scale toward and around lines of sight to BN 
and is consistent with a twist of the magnetic field on a larger scale towards OMC-1. 
A comparison of our results with several other studies suggests that a two-component magnetic field, 
maybe helical, could be wrapping the OMC-1 filament. 
\end{abstract}


\newpage

\section{INTRODUCTION} \label{INT}

In order to better understand the nature of magnetic fields observed in the diffuse 
interstellar medium (ISM) and in higher density regions of the ISM, we compare results 
from two observational techniques, visible and near-infrared (NIR) polarimetry 
of background stars and submillimeter (submm) polarimetry in dark clouds. 
Such a comparison between two different wavelength regimes 
was applied in a previous paper by \citet{poi106} to the GF 9 region.
Here it is applied to the filamentary star-forming molecular clouds in the Orion A region.  

Visible polarimetry by pioneers such as \citet{hil49} and \citet{hal49}, followed 
by others such as \citet{mat70}, led to a first picture of the Galactic
magnetic field assuming that dust grains pervading the diffuse ISM are magnetically aligned
with their long axis preferentially perpendicular to the local magnetic field.
In this context, polarization is produced by dichroic extinction.
The second technique, polarimetry at longer wavelengths, successfully investigated in the far-infrared (FIR) 
by \citet{cud82}, and in the submm by \citet{hil84}, allowed the first detections of the   
polarized radiation emitted by aligned dust grains in several cold dense regions in the ISM
including OMC-1, thus probing magnetic fields in these environments. 
In parallel to improvements in both techniques, a general dust grain alignment 
theory was developed and is now arriving to maturity as reviewed by \citet{laz03}
and \citet{laz07}. 

Since the advent of both techniques, several observations focused on the Orion 
region. The ``Integral Shaped Filament'' (ISF) observed at submm wavelengths 
by \citet{joh99} is part of the molecular clouds located in the 
Barnard's/Eridanus loop, 
a diffuse and expanding shell of enhanced optical emission covering an area of 
$\approx 15^{\circ} \times 8^{\circ}$. Located to the south of Orion B and 
to the north of the L1641 dark nebulae, the ISF lies in the northern portion 
of Orion A which is one of the most active sites of star formation in the solar vicinity.
It is associated with the well-known Orion nebula and the bright Trapezium stars. It also contains the OMC-1
cloud core located behind the nebula, the two extensively studied star-forming filaments,
OMC-2 and OMC-3, located $\approx 15'$ and $25'$ to the north \citep[see][]{cas95}, 
IRAS 05327-0457 located $\approx 30'$ to the north and OMC-4 located 
$\approx 8'$ to the south \citep[see][]{hou04} of OMC-1. 
 
A combination of 21 cm emission-line Zeeman splitting observations with 
a compilation of visible polarimetry by \citet{hei97} gives a relatively complex picture 
of the magnetic field on large spatial scales in the Eridanus/Orion region. 
On smaller spatial scales, visible polarimetry observations were carried out by \citet{bre76}
and \citet{bre77} in the vicinity of Orion A. Based on these data the mean magnetic 
field projected on the plane-of-the-sky (POS) appears to be consistent with 
the picture of the magnetic field depicted within the clouds by \citet{sch98} from 
$100$ $\mu$m and $350$ $\mu$m polarimetry in OMC-1. Polarimetry at 
$\lambda=850$ $\mu$m by \citet{mw00} and \citet{mwf01} shows
that a helical magnetic field could thread the northern region of the 
OMC-3 dust filament. Additional polarimetry was obtained by \citet{hou04} 
at $\lambda=350$ $\mu$m in OMC-2 and OMC-3 and by \citet{poi206} at $850$ $\mu$m 
in OMC-2. Comparison of these data revealed that the polarization patterns are 
fairly similar at both wavelengths. 
\citet{poi206} found that uniform magnetic field components are dominant compared to turbulent ones
in OMC-3, but uniform and turbulent components are comparable in OMC-2. 

The work proposed in the following is partially complementary to the one recently 
presented by \citet{li09} where a comparison of the magnetic fields probed at 
scales between the accumulation length and the size of cloud cores is discussed.
\citet{li09} show that for field directions detected at the two extremes, a significant 
correlation is found. On the contrary \citet{poi106} find that the large scale uniform magnetic field 
observed in the vicinity of the GF-9 filaments could be dragged on smaller scale in one of the rotating cores. 
In order to improve our knowledge of the morphology of the 
magnetic fields on intermediate scales and its interpretation, new visible and NIR 
polarization observations are presented. These observations cover the Orion A region 
and its vicinity and compensate for the previous under-sampling of the magnetic 
field on scales of a few times the width of the ISF
seen on submm emission maps \citep{joh99}. This corresponds to a scale of $\approx$
2.5 pc on the POS for a distance to the Nebula of about 414 pc \citep[see][]{men07}.

The new data with a compilation of published visible polarization
data are presented in section \ref{OBSRES}. 
Submm, visible and NIR data are used in section \ref{comparisons} to compare the 
direction of magnetic fields projected on the POS
as probed with the two techniques on spatial scales ranging from $\approx 2^{\circ}$ to $\approx 1'$. 
This section is structured as follows: first a comparison of results on large 
spatial scales ($\approx 1^{\circ} \times 4^{\circ}$) is presented. 
Then regions OMC-1, OMC-2/3 and OMC-4 are discussed and compared to fields from 
large to small spatial scales when appropriate. NIR data in the Becklin-Neugebauer 
(BN) region are extensively analyzed and compared with submm polarimetry data. 
A discussion of the results follows in section \ref{discussion} and a summary in section \ref{sum}.     

\section{OBSERVATIONAL RESULTS} \label{OBSRES}

\subsection{Visible Polarization Data in and around Orion A}

\subsubsection{Observations} \label{OBS}

Observations were carried out on the 1.6 m Ritchey-Chr\'etien telescope of 
the Observatoire du Mont-M\'egantic (OMM), Qu\'ebec, Canada,
between 2003 January and 2004 November with Beauty and the Beast, a two-channel 
photoelectric polarimeter, which uses a Wollaston prism, a Pockels cell, and 
a quarter wave plate with a $8\farcs2$ aperture hole and a broad red
filter (RG645: 7660 {\AA} central wavelength, 2410 {\AA} FWHM). The data 
were calibrated for instrumental efficiency, instrumental polarization (due to the telescope
mirrors) and zero point of the position angles using a calibration prism, 
non polarized standard stars, and polarized 
standard stars, respectively. The instrumental polarization was measured to be $0.054 \pm 
0.068 \%$ and was subtracted from the observed polarization values. 
The observational errors were calculated from photon statistics 
and also include uncertainties introduced by the previously mentioned calibrations.
The final uncertainty on individual measurements of the polarization $P$ is 
usually around $0.1 \%$. The data were de-biased in the standard way \citep[e.g.][]{war74}.
More details about the instrument and the observational 
method can be found in \citet{man95}.   

Additional polarimetry was obtained on 14 sources in OMC-2
during 1992 October at the Mount Lemmon Observing 
Facility (MLOF) 60$\arcsec$ using the Two-Holer Photopolarimeter with Cousins-Kron $V$, $R$ and $I$ filters. 
A description of this instrument can be found in \citet{sit85}.  
    
IR polarimetry on stars in OMC-1 near the BN object is described 
below in section \ref{BNregion}. 

\subsubsection{Compilation of the Data} \label{datavis}

A set of 69 stars located throughout the whole Orion A region 
was observed at OMM. Among this set, 41 objects had never been  
observed before in polarimetry; these new data are presented in Table \ref{OMMDATA}.
Table \ref{PVISDATA} presents data on stars observed more than once, at OMM
and published by \citet{bre76} or found in the \citet{hei00} catalog. 
Table \ref{OTHERDATA} presents data on other stars published by \citet{bre76}
or in the \citet{hei00} catalog. Following \citet{bre76} the polarization 
mainly produced either by ISM dust or by Intracluster Dust (ICD) is indicated in the Tables.
The effective wavelength of the stars observed 
by \citet{bre76} lies between that of the $B$ and $V$ filters. On the other hand, wavelengths at which stars 
were observed are not specified in the \citet{hei00} catalog
but in many cases the $V$ filter was used.
In each Table, Parenago and when available HD numbers for each star 
are shown in columns 1 and 2, respectively. 
Equatorial coordinates at epoch J2000.0 are given in columns 3 and 4. 
Measured polarization values with uncertainties, and position angles 
with their uncertainties are given in columns 5 to 8, respectively. 
Then `1' is tabulated in column 9 if the signal to noise ratio 
$P / \sigma_{P} > 3$ (meaning $\sigma_{\theta} < 9\fdg5$)
otherwise `0' is tabulated. Column 10 gives an approximate value 
of the visible absorption coefficient $A_{\rm V}$ from the 
\citet{dob05} catalog. This catalog is the first version of the atlas and 
catalog of dark clouds derived by using an optical star-count technique on the 1043 
plates in the DSS. Distance estimates retrieved from the \citet{hei00} catalog 
are shown in column 11. 
 
Polarization data for the stars observed from Mount Lemmon are compiled in Table \ref{OMC2DATA}.
The identification number appearing in the first column is from \citet{jon94} 
where $JHK$ photometry data are compiled in their Table 1. 
J2000.0 coordinates are given in columns 2 and 3. 
Results of the polarimetry in the three bands are given in columns 4 to 15.   
The extinction coefficient appearing in column 16 is from the \citet{dob05} catalog. 

\subsubsection{Classification of the Data} \label{pvisclass}

A quality code (SC) is shown in the last column of Tables \ref{OMMDATA}, \ref{PVISDATA} and 
\ref{OTHERDATA}. The selection criteria used to classify the data, and to distinguish
polarization produced by scattering on dust grains from polarization produced 
by dichroic extinction, are: `1' is tabulated for data for 
which the polarization is produced 
by dichroic extinction with great reliability, meaning that (1) the position angles  
do not rotate more than $20^{\circ}$ when observations are available at several 
wavelengths, or that (2) no variability of the degree of polarization and of the position angle 
can be found with time at a given wavelength. Some of the stars listed in Tables 
\ref{PVISDATA} or \ref{OTHERDATA} 
are marked with `1' following results from \citet{bre76} and \citet{bre77}. 
The value `2' is tabulated for (3) data with a position angle approximately the same as 
those of reliable data in their vicinity. `?' is tabulated for data for which we do not have
enough information in light of the three conditions mentioned above, and `0' is tabulated 
for data in contradiction with conditions (1) and (2). 
Finally, `0' is systematically tabulated when the signal to noise ratio $P / \sigma_{P} < 3$.        
Following the analysis discussed 
in the next section (\ref{pvisext}), a few other stars also have their quality code forced to SC $=0$. 

The selection code shown in the last column of Table \ref{OMC2DATA} follows the same conventions 
except that data with $P/\sigma_{P} < 3$ (meaning $\sigma_{\theta} > 9\fdg5$) are also 
positively considered here since their signal-to-noise ratio is never far from 
the ratio $P/\sigma_{P} = 3$.  

\subsubsection{Distribution of Visible Polarization with Visual Extinction} \label{pvisext}

\begin{figure}
\epsscale{0.8}
\plotone{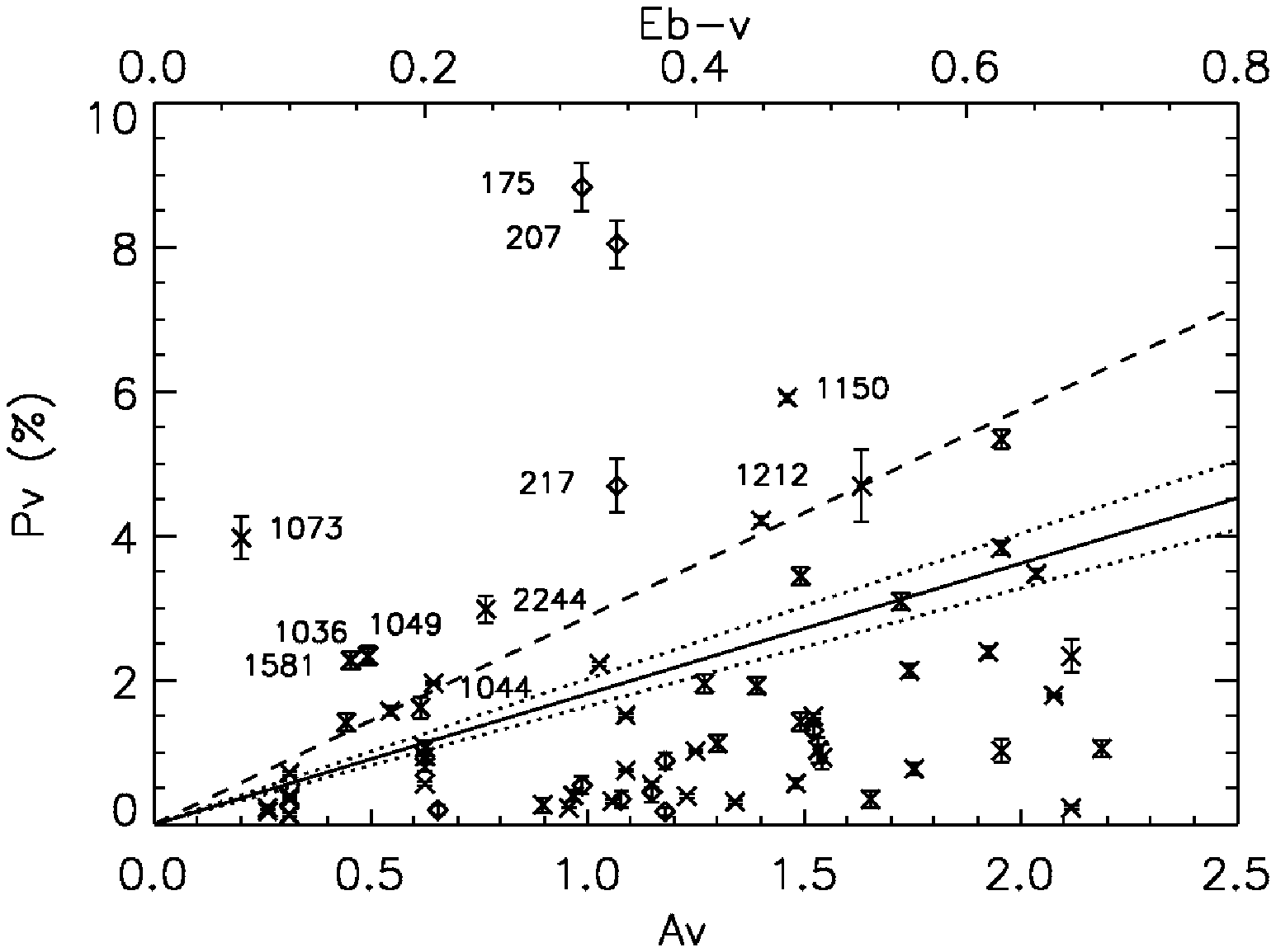}
\caption{Variations of $P$ with $A_{\rm V}$.
Upper axis shows $E_{\rm{B-V}} = A_{\rm V} /3.1$ for comparison. 
The mean of selective to total extinction ratios $\overline{R} = 4.93 \pm 0.51$ from 
\citet{duc03} is used. The solid line shows the limit $P = 9 E_{\rm{B-V}} $ when
parameter $\overline{R} = 4.93$ is used. Dotted lines show the corresponding 
limits covered by this relation when uncertainties on $\overline{R}$ are included.
The dashed line shows the same limit when the usual value of  
$\overline{R} = 3.10$ is considered. The reliability of the data refers to their
reliability to probe magnetic fields (see section \ref{pvisclass} for details).  
Crosses show reliable and less reliable data from Tables \ref{OMMDATA}, \ref{PVISDATA}
and \ref{OTHERDATA}. Diamonds
also show reliable and less reliable data from Table \ref{OMC2DATA} 
where the $R$-band is used.   
\label{pav}}
\end{figure}

The distribution of polarization, $P$, with visual extinction, $A_{\rm V}$, is shown in 
Figure \ref{pav}. 
The upper axis shows $E_{\rm{B-V}}= A_{\rm V}/R_{\rm V}(= 3.1)$. 
The dashed line shows the limit \citep{ser75} $P = 9 E_{\rm{B-V}} = 2.9 A_{\rm V}$ with the usual 
parameter $R = 3.1$. However, the mean of selective to total extinction ratios has been 
determined in Orion A by \citet{duc03} to be $\overline{R} = 4.93 \pm 0.51$. 
The full line shows the limit $P = 9 E_{\rm{B-V}}$ when  
$\overline{R} = 4.93$ is used instead of the normal interstellar value. 
Dotted lines show the corresponding limits when uncertainties on $\overline{R}$ are included.
Crosses show reliable and less reliable data from Tables \ref{OMMDATA}, \ref{PVISDATA} and 
\ref{OTHERDATA} and diamonds show reliable and less reliable data from Table \ref{OMC2DATA} where 
the $R$-filter data is used.
We recall the $A_{\rm v}$ values come from \citet{dob05}, are based on an optical star-count technique
and therefore could be good estimators of the visible absorption along the 
LOSs to the observed stars or upper limits on the visible absorption along the LOSs 
to stars foreground to or poorly embedded into the cloud.

For this reason they do not depend on the value of $R_{\rm V}$ one may adopt. 
On the other hand, the value of $R_{\rm V}$ one may use gives upper limits on $E_{\rm{B-V}}$. 
In the following we adopt the usual value of $R_{\rm V} = 3.1$. 
We define $E_{\rm{B-V, lim}}= A_{\rm V}/3.1$ and reject data with
$P_{\rm V} > 9 E_{\rm{B-V,lim}}(= 2.9 A_{\rm{V}})$.   

Many points can be seen in Figure \ref{pav} above the upper limit $P = 2.9 A_{\rm{V}}$. 
Some of these objects are stars observed by \citet{bre76} and references therein; 
their well characterized polarization measured at several epochs in several bands is produced 
by dichroic extinction by intra cluster dust (ICD) or by the foreground ISM according to his analysis.
Therefore we chose to include it in our analysis.
Objects not discussed by \citet{bre76} are star P2244 and IR sources 175, 207 and 
217 \citep[see Table \ref{OMC2DATA} and][]{jon94}. 
Multiwavelength observations shown in Table \ref{OMC2DATA} suggest that polarization of objects 
175 and 207 could be produced by dichroic absorption mechanisms; this would need to be confirmed 
by further observations and we chose to exclude these objects from our analysis. 
Stars P2244 and IR source 217 were also removed from the sample.

In conclusion, all visible data displayed in Tables \ref{OMMDATA}, 
\ref{PVISDATA}, \ref{OTHERDATA} and \ref{OMC2DATA} are classified as follows:
data defined as reliable to probe magnetic fields are data for which 
the selection code `1' or `2' is tabulated.
Data considered as less reliable are marked with `?'. 
Data considered as unreliable are marked with `0'. 
In our analysis only reliable and less reliable data for probing magnetic fields are considered. 
In the rest of the paper, unless stated otherwise, 
the term reliable will be used to refer to the reliability with regards to probing 
magnetic fields as gauged by the selection code.

\subsubsection{Visible Polarization Map and Histograms of Polarization Data} \label{polhistvis}

\begin{figure}
\epsscale{1.}
\plotone{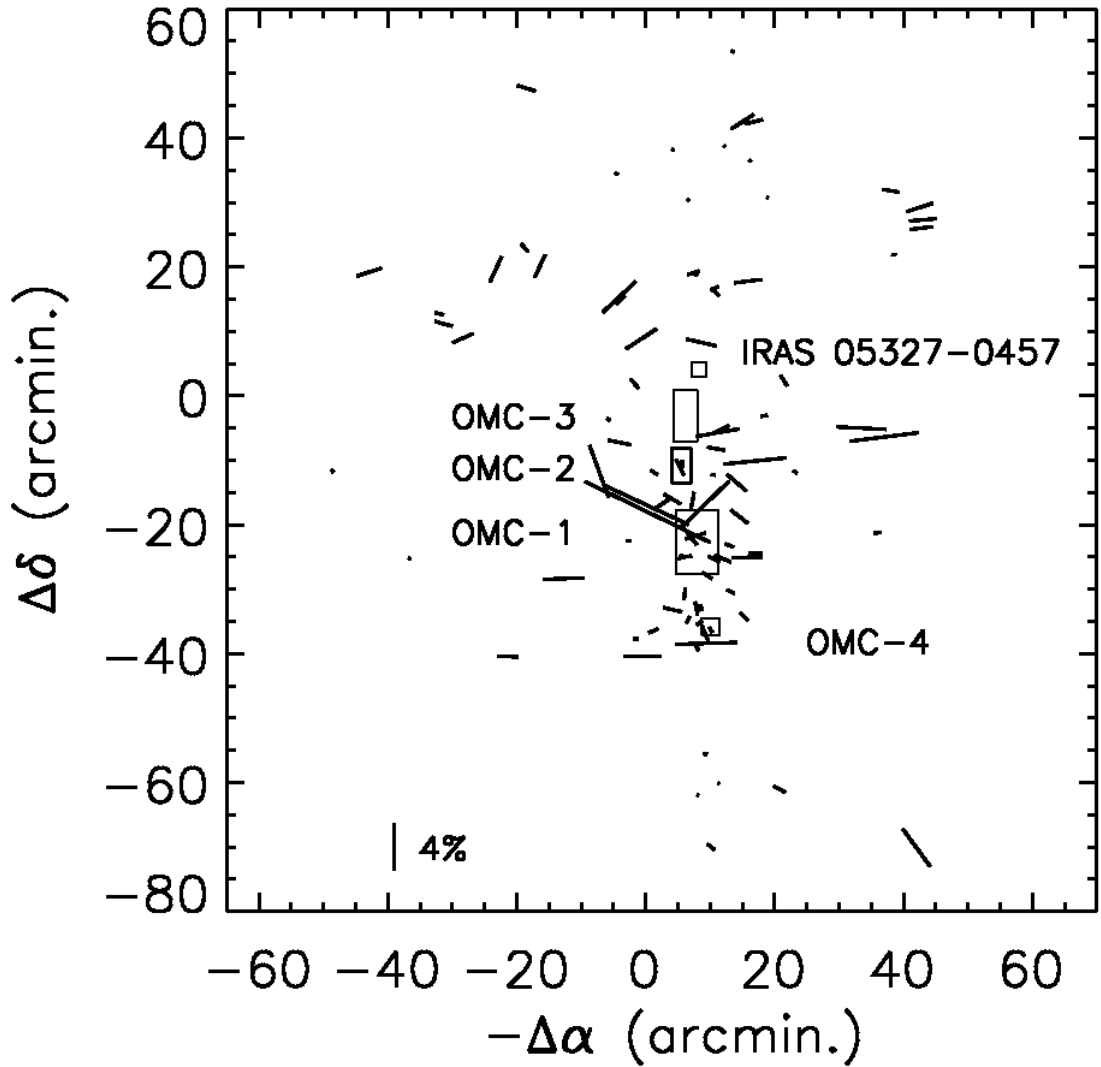}
\caption{Visible polarization map of Orion A. The reliability of the data refers to their
reliability to probe magnetic fields (see section \ref{pvisclass} and \ref{pvisext} for details). 
Reliable data labeled with `1' or `2' and less reliable data labeled with `?' 
from Tables \ref{OMMDATA}, \ref{PVISDATA} and \ref{OTHERDATA} are shown in the Figure. 
Five different regions identified by boxes are labeled.
The reference position is R.A.=5$^{\rm h}$35$^{\rm m}$48.0$^{\rm s}$, decl.=-5$^{\circ}$ 00$^{\rm mn}$ 00.0$^{\rm s}$ (J2000.0). 
See Figure \ref{lsunp} for a better polarization position angle visualization.
\label{cartevis}}
\end{figure}

\begin{figure}
\epsscale{0.8}
\plotone{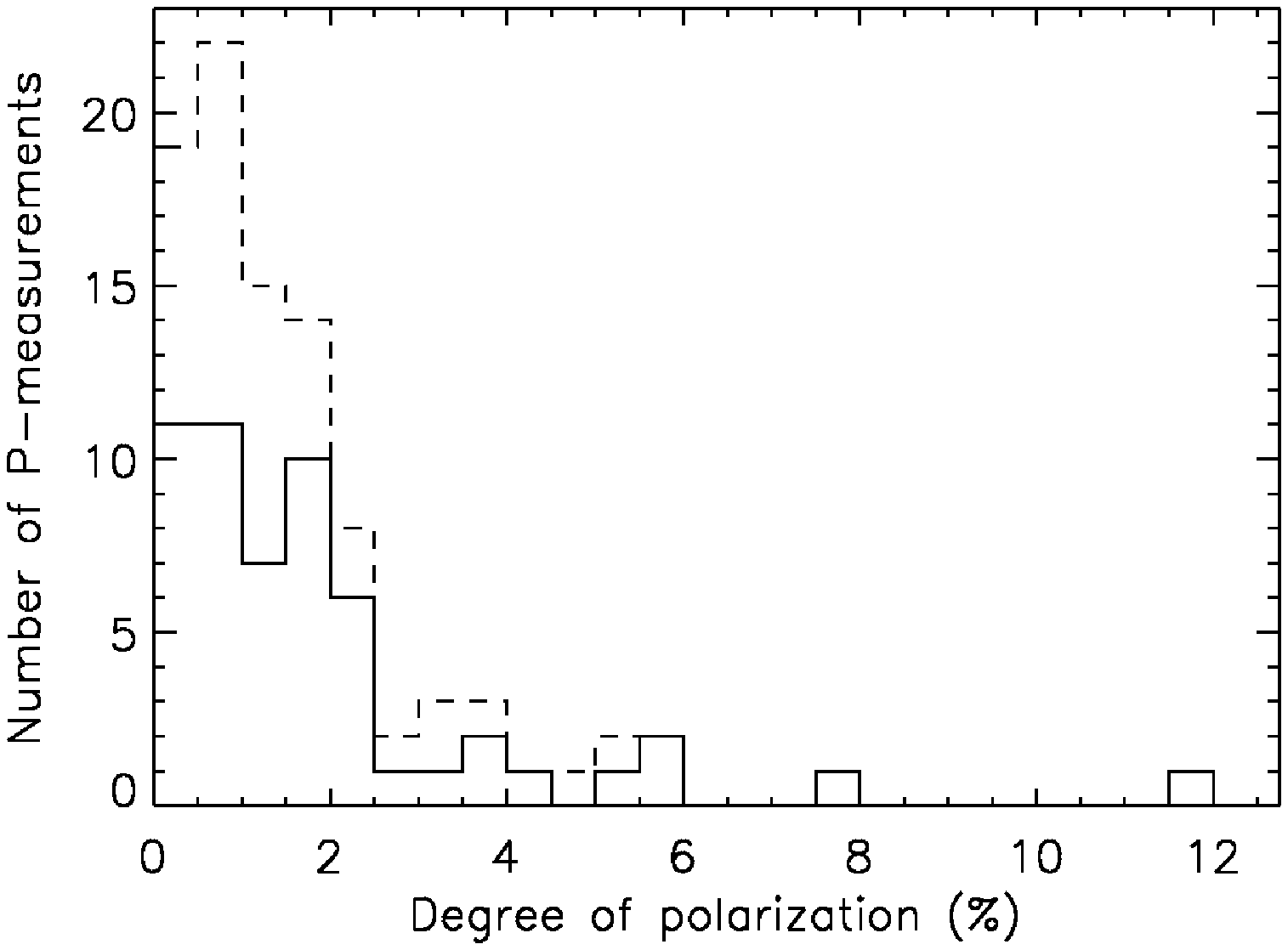}
\caption{Histogram of the degree of polarization of visible data 
from Tables \ref{OMMDATA}, \ref{PVISDATA} and \ref{OTHERDATA} shown in Figure \ref{cartevis}. 
Data considered as reliable data to probe magnetic fields (SC = 1 or 2) are shown with full lines. 
Less reliable data (SC = ?) are shown with dashed lines. 
\label{phisto}}
\end{figure}

\begin{figure}
\epsscale{0.8}
\plotone{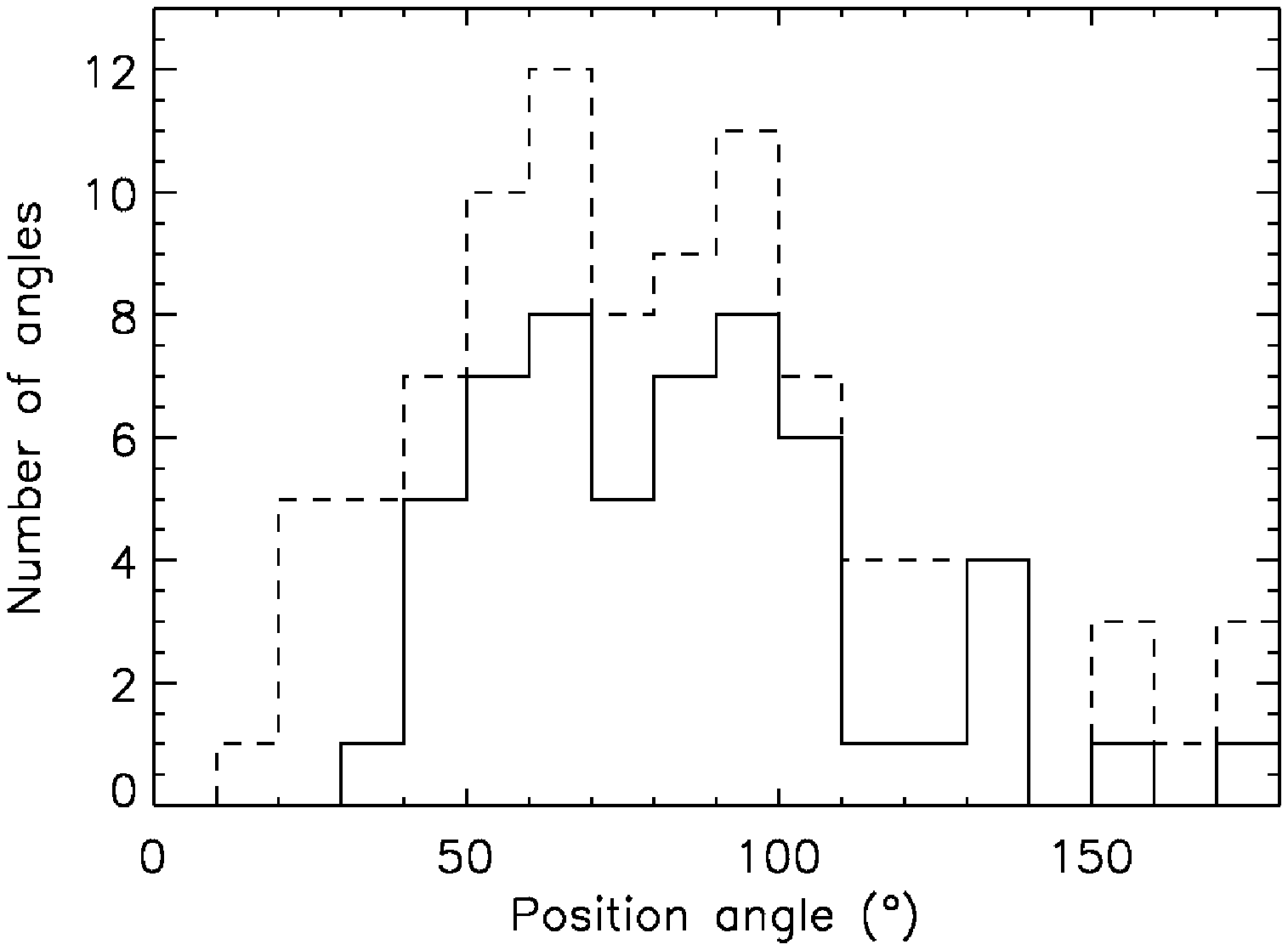}
\caption{
Same as in Figure \ref{phisto} but for position angles.
\label{anghisto}}
\end{figure}

A polarization map of reliable and less reliable data from Tables \ref{OMMDATA}, 
\ref{PVISDATA} or \ref{OTHERDATA} is shown in Figure \ref{cartevis}. 
When several measurements are available (see Tables 
\ref {PVISDATA} and \ref{OTHERDATA}), data with the best signal-to-noise ratio
are shown in the Figure. The data cover an area of $\approx 1\fdg5 \times 2\fdg1$
centered on the region OMC-2, meaning an area of about 15 pc $\times$ 21 pc for a 
distance to the Nebula of about 414 pc \citep[see][]{men07}. Histograms of polarization 
percentages and position angles of the data shown in Figure \ref{cartevis}
are presented in Figures \ref{phisto} and \ref{anghisto}, respectively.
In Figure \ref{phisto}, the mean and the dispersion of the whole data set ($N$=94)
are $\overline{P} = 1.70\% $ and $S_{\rm P} = 1.81 \%$, respectively. Reliable data ($N$=55)
are shown with full lines and the mean and the dispersion of this data set are 
$\overline{P} = 1.92\% $ and $S_{\rm P} = 2.12 \%$, respectively. 
The polarization position angle is a variable that wraps on over itself therefore the means 
retained in our analysis correspond to the means obtained where the dispersions 
of the distributions are found to be the smallest. 
In Figure \ref{anghisto}, the mean and the dispersion of the whole data set
are $\overline{\theta} = 75\fdg0$ and $S_{\theta} = 36\fdg5$, respectively.
Reliable data are shown with full lines and the mean and the dispersion of  
this data set are $\overline{\theta} = 78\fdg5$ 
and $S_{\theta} = 29\fdg8$, respectively.

\begin{figure}
\epsscale{1.0}
\plotone{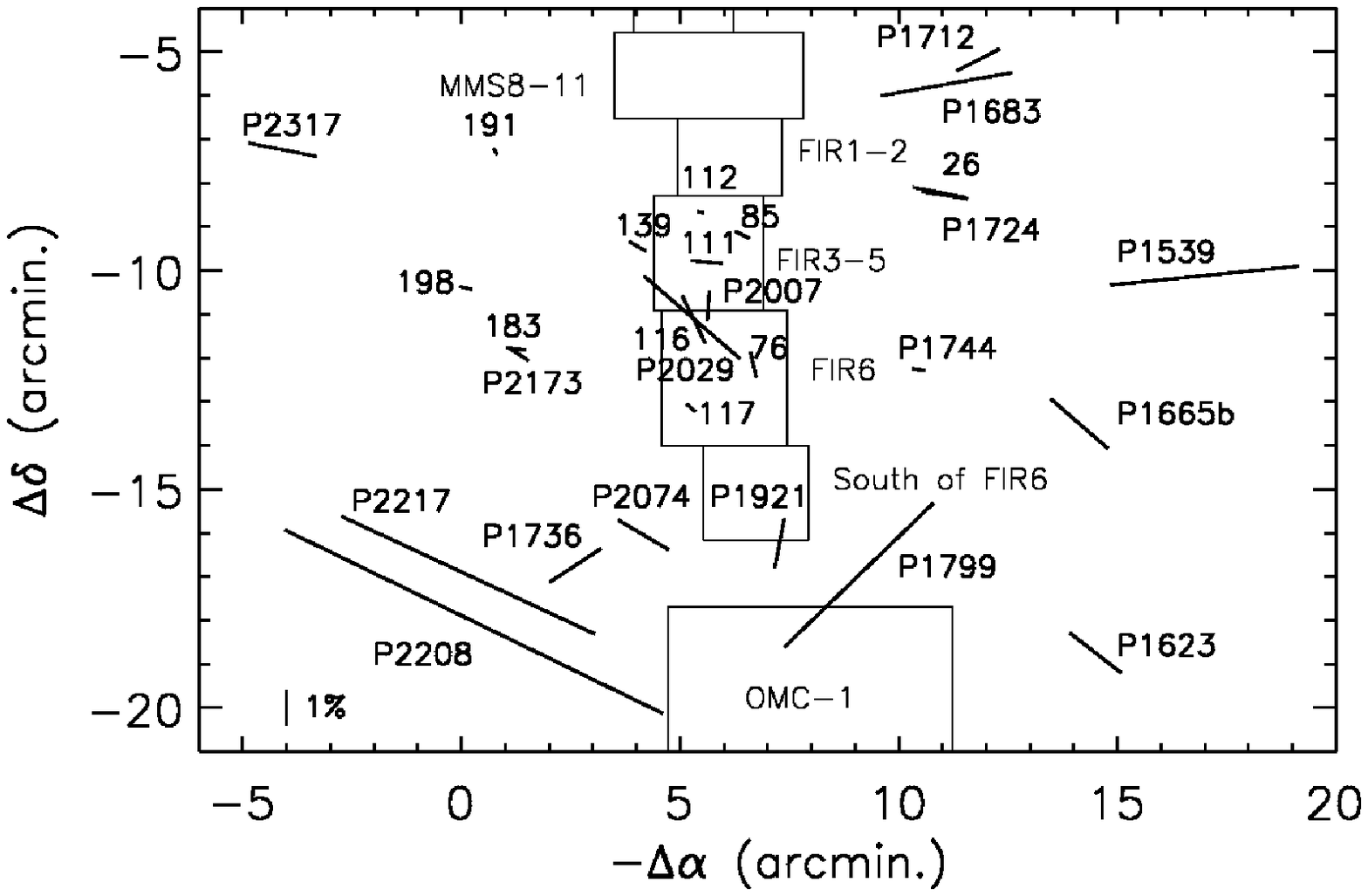}
\caption{Same as in Figure \ref{cartevis} but with a zoom on OMC-2. 
Selected $R$-band data from Table \ref{OMC2DATA} are now included. 
Subregions of OMC-2, FIR1-2, FIR3-5, FIR6, and south of FIR6 are indicated by boxes. 
The reference position is R.A.=5$^{\rm h}$35$^{\rm m}$48.0$^{\rm s}$, decl.=-5$^{\circ}$ 
00$^{\rm mn}$ 00.0$^{\rm s}$ (J2000.0). 
\label{carteomc2}}
\end{figure}

\begin{figure}
\epsscale{0.8}
\plotone{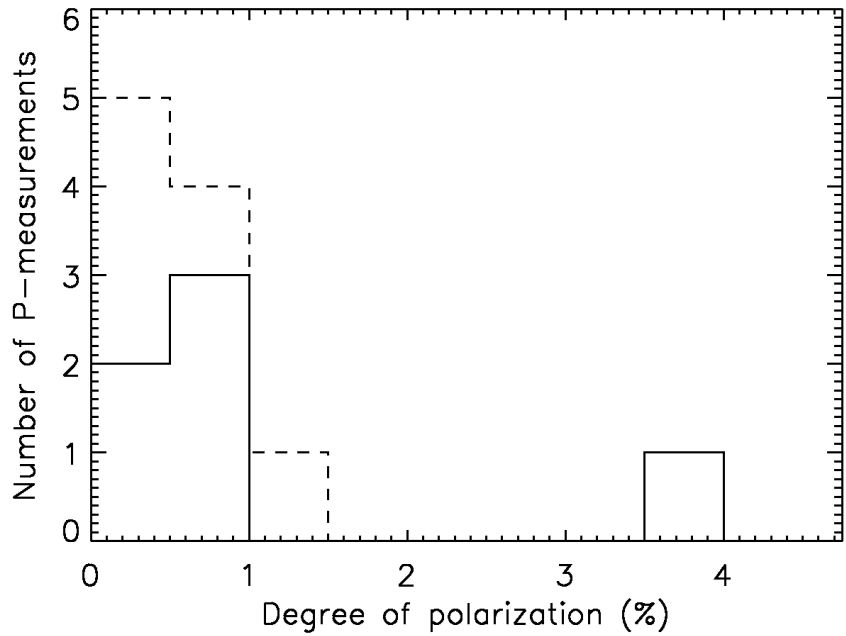}
\caption{ Histogram of the degree of polarization of the selected $R$-band data from Table \ref{OMC2DATA}
shown in Figure \ref{carteomc2}. Data considered as reliable data to probe magnetic 
fields (SC = 1 or 2) are shown with full lines. Less reliable data (SC = ?) are shown with dashed lines. 
\label{phisto2}}
\end{figure}

\begin{figure}
\epsscale{0.8}
\plotone{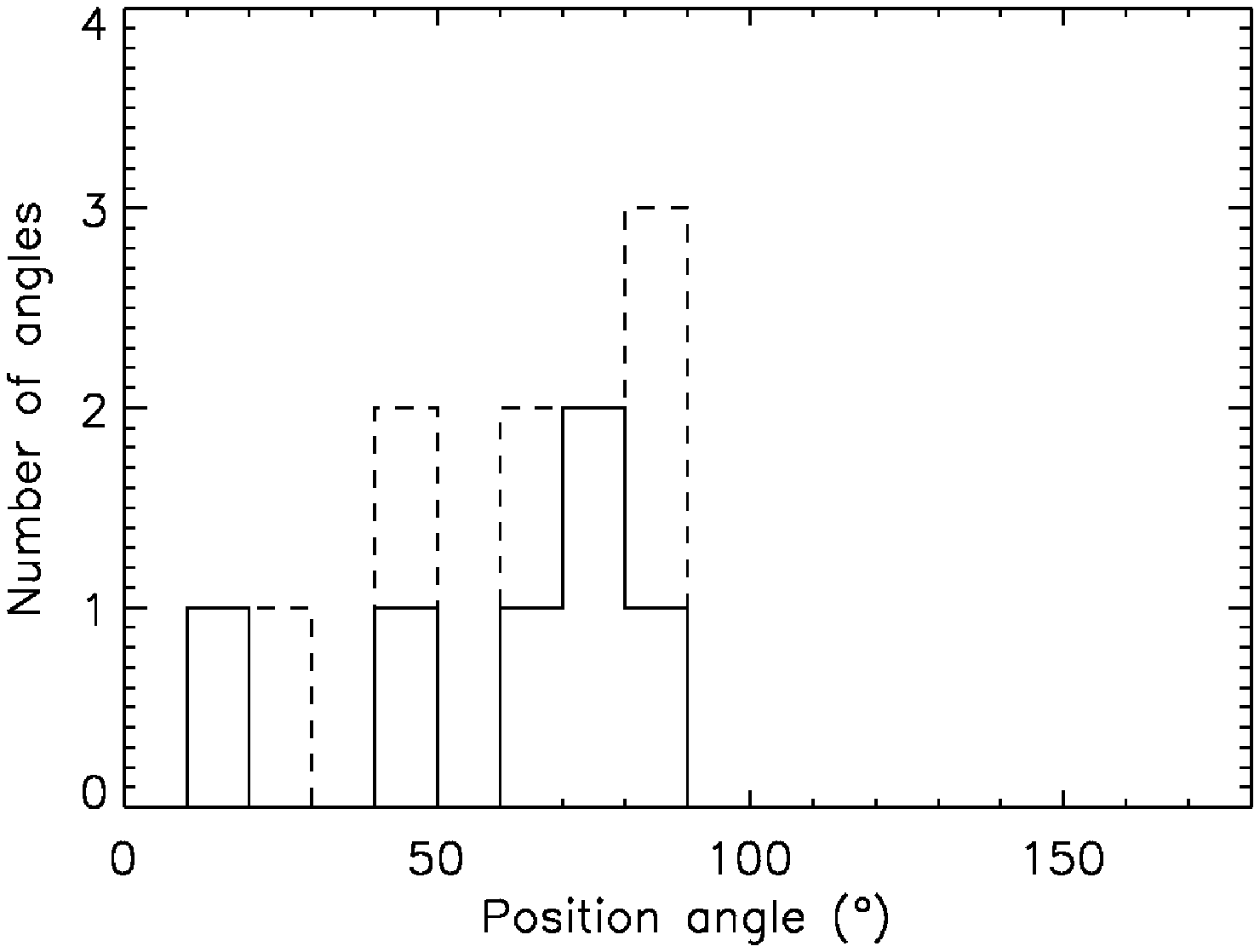}
\caption{ 
Same as in Figure \ref{phisto2} but for position angles.
\label{anghisto2}}
\end{figure}

Reliable and less reliable data from Table \ref{OMC2DATA} are shown in Figure \ref{carteomc2}. 
Also shown in the map are the data selected from Tables \ref{OMMDATA}, \ref{PVISDATA} and 
\ref{OTHERDATA} located in the region. Data from Table \ref{OMC2DATA} cover an area of 
$\approx 0\fdg2 \times 0\fdg1$ (or about 2 pc $\times$ 1 pc for a 
distance of about 414 pc to the Nebula \citep[see][]{men07}) with six stars located in the 
LOSs to regions FIR3 to FIR6 \citep[see][]{chi97}. Stars P2007 and P2029 from 
Table \ref{OMMDATA} are also located in similar LOSs. Following \citet{jon94}, 
all the stars from Table \ref{OMC2DATA} are brighter than $K=+13$ and are not field stars. 
They must be members of a young association or associations within the observed region.   
Histograms of polarization percentages and position angles 
of reliable and less reliable data from Table \ref{OMC2DATA}
are shown in Figures \ref{phisto2} and \ref{anghisto2}, respectively. 
In Figure \ref{phisto2}, the mean 
and the dispersion of the whole data set ($N$=11) 
are $\overline{P} = 0.82\% $ and $S_{\rm P} = 0.97 \%$, respectively. 
Reliable data only ($N$=6) are shown with full lines and their mean and dispersion  
are $\overline{P} = 1.04\% $ and $S_{\rm P} = 1.26 \%$, respectively. 
In Figure \ref{anghisto2},
the mean and the dispersion of the whole set of data ($N$=11)
are $\overline{\theta} = 60\fdg9 $ and $S_{\theta} = 23\fdg6$, respectively.
Reliable data are shown with full lines ($N$=6) and their mean and dispersion   
are $\overline{\theta} = 60\fdg7$ and 
$S_{\theta} = 26\fdg8$, respectively.

\subsection{NIR Data in the Direction of the BN region} \label{BNregion}

\subsubsection{Observations} \label{databn}

Polarimetry on and around the BN object 
was done on 33 stars on 2002 February 15 (UT) using NSFCAM on the 
IRTF in polarimetry mode. The filter was the standard IRTF $K$-band filter centered on
$\lambda = 2.2$ $\mu$m. A description of NSFCAM in polarimetry mode was given 
by \citet{jon97}. An identification number, J2000.0 coordinates, 
the degree of polarization, the position angle and their uncertainties 
are given in columns 1 to 6 in Table \ref{BNDATA}, respectively. Color indices $H - K$ 
are given in column 7. A precision of $\pm 0.3 \%$ or better was reached for each observation 
and the relation, $\sigma_{\theta_{K}} = 28\fdg6 / (P_{K}/\sigma_{P_{K}})$, was used 
with $\sigma_{P_{K}} = 0.3 \%$, to estimate $\sigma_{\theta}$ in column 6.
The optical depth $\tau_{K}$ at $\lambda = 2.2$ $\mu$m was estimated for each of the 33 sources
with equation (1) from \citet{jon89} by using the relation $\tau_{K} = 1.5 E_{H-K}$.
The color excess was estimated by subtracting the intrinsic color index of middle 
M red dwarfs \citep{lad04} $(H-K)_{0}=0.3$ from $H-K$ (column (7) of Table \ref{BNDATA}). 
Estimates of $\tau_{K}$ are displayed in column 12 of Table \ref{BNDATA}. 
The use of $(H-K)_{0}=0.3$ might introduce a bias in the results as one can see in 
Table \ref{BNDATA} that negative values of optical depth are found for stars number 5 and 35.
Since the median value of the distribution of $H-K$ is 0.83 we believe, however, that most 
of our estimates are good approximations of the optical depth.   

Since coordinates of each object were defined relative to a zero coordinate origin
during the observations, coordinates from BN and a value of $\approx 0\farcs9/$pixel 
corresponding to the plate scale were used to compute the coordinates given in columns 2 and 3.  
A search for a possible identification of each object was then made by using the 
Simbad database. In many cases several objects are located a few arcseconds from these 
positions and each of the sources compiled in Table \ref{BNDATA} can generally 
be identified as or associated to an IR source (IR), a Part of Cloud (PoC) source, a Maser (Mas), 
a Star in Cluster (*iC), a Star in Nebula (*iN), a Parenago object or a X-ray source (X). 
These possibilities are not all included in Table \ref{BNDATA}, but by comparisons with  
works by \citet{lon82} (LBLS) and \citet{sto98}, possible identifications for some of the sources 
are given in columns 8 to 10 in Table \ref{BNDATA}. Sources 18, 27 and 31 
are suspected to be identified to objects named CB1, CB4 and n respectively and discussed 
by \citet{sto98} and/or by \citet{shu04} (S$\&$S). CB1 and CB4 are Close Binaries while n 
is a point-like source and is believed to be surrounded by a circumstellar disk. Estimates of 
$\theta_{\rm off}$ displayed in columns 11 will be introduced 
and discussed in the following sections.        

Polarimetry on BN is consistent with $\lambda = 2.2$ $\mu$m polarimetry by \citet{lon80} 
and \citet{min91}, and with $\lambda = 3.8$ $\mu$m polarimetry by \citet{dou93} on this source.
However $\lambda=2.2$ $\mu$m polarimetry on source 31 (n) differs from $\lambda = 3.8$ $\mu$m 
polarimetry by \citet{dou93} on this source where $P \pm \sigma_{P} = 16.6 \pm 1.2 \%,
\theta \pm \sigma_{\theta} = 110 \pm 2 ^{\circ}$ with an angular resolution of $0\farcs8$, and where  
$P \pm \sigma_{P} = 12.6 \pm 4.0 \%, \theta \pm \sigma_{\theta} = 115 
\pm 8 ^{\circ}$ with an angular resolution of $0\farcs2$. A comparison with the $\lambda = 2.2$ $\mu$m
map by \citet{min91} shows that the polarization pattern shown in our map differs greatly 
from their centro-symmetric polarization vector pattern. Some of the vectors in our 
map appear sometimes to have a similar orientation to some vectors in their map
but the comparison is difficult and limited without a tabulated version of their data.
An outflow/disk system around BN was revealed with $K$-band high resolution 
observations by \citet{jia05}.
In addition, $\lambda = 2.2$ $\mu$m circular polarization of OMC-1 by \citet{bus05},
wide-field and deep $JHK_{s}$-bands polarization images of the Orion Nebula by \citet{tam06},    
HST NICMOS polarization measurements of OMC-1 at $\lambda = 2.2 $ $\mu$m with $0\farcs2$
resolution by \citet{sim06}, $JHKs$-band polarimetry of $\approx 500$ stars of the Orion Nebula Cluster 
in M42 by \citet{kus08}, and $K_{s}$-band circular and linear polarization by \citet{fuk09} are available 
on and around BN. A multi-data survey was conducted in OMC-1 by \citet{val07} and several
configurations of magnetic fields are considered and discussed.
The four possible mechanisms proposed for explaining circular polarization observed on large scale 
in OMC-1 are discussed and some of them investigated by \citet{mats09}.
Our data will be discussed and compared with some of these works.
 
\subsubsection{Selection of the Data}

\begin{figure}
\epsscale{0.8}
\plotone{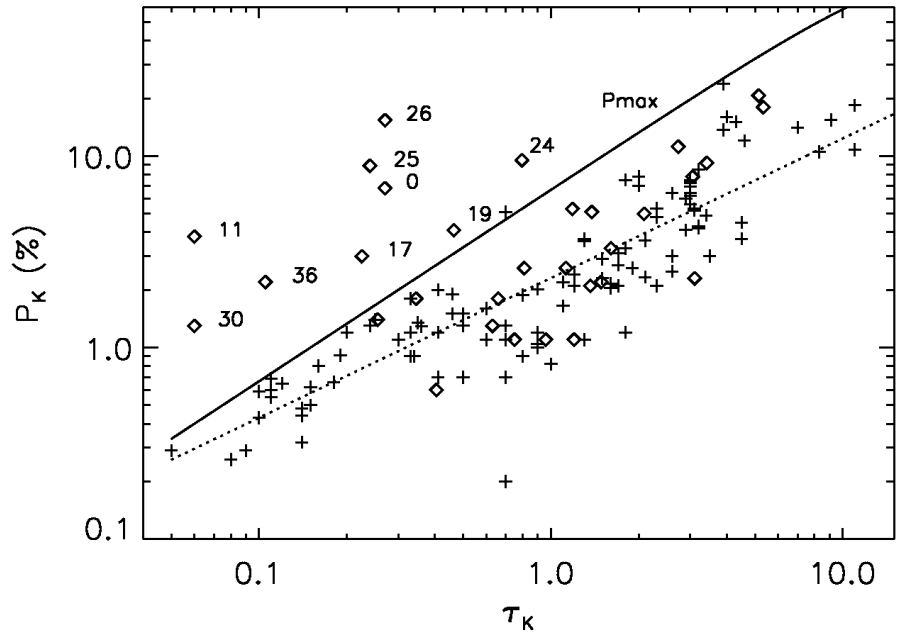}
\caption{Variations of $P_{K}$ with $\tau_{K}$. Crosses show the distribution of 
the data discussed by \citet{jon89}. Diamonds show the distribution of the new data 
obtained in OMC-1. The dashed line shows the fit to all the data shown in the Figure 
once those lying above the $P_{\rm max}$ line are removed from the sample (see text for details). 
\label{pvstau}} 

\end{figure}

The statistical behaviour of polarization efficiency at 2.2 $\mu$m has been explored by \citet{jon89}.
A good correlation is found between interstellar extinction and interstellar polarization over a range in optical
depth of a factor of 100. A least-squares fit to the sample of data obtained on 105 sources has a slope of 0.75
corresponding to the relation, $P_{K} = 2.23 \tau_{K}^{3/4}$. The distribution of these data 
is shown with crosses in Figure \ref{pvstau} \citep[see also][]{jon89}. 
Except for stars number 5 and 35 toward which the estimates of $\tau_{K}$ are negative, 
the distribution of the new data set in OMC-1 is shown with diamonds, for comparison. 
The upper solid line in Figure \ref{pvstau} corresponds 
to equation (A7) from \citet{jon89}, where $\eta = 0.875$ is used. With this choice 
of $\eta$, the solid line displayed in the Figure is equivalent to the one 
displayed in our Figure \ref{pav} and translates 
the relation $P_{\rm max} = 9.0 E_{B-V}$ estimated by \citet{ser75}.   

We see that nine sources show a polarization degree higher than the upper limit. 
These stars are identified in the Figure and, in addition to sources 5 and 35, they are excluded 
from the analysis since their polarization could be contaminated by intrinsic polarization due to scattering
as illustrated by \citet{kus08} for some sources of their $JHK_{\rm s}$-band polarization sample. 
The quality code SC associated to the rejected measurements is SC $=0$. 
In a complementary way, the data located below the upper limit have a quality code SC $=1$. 
The quality code is displayed in the last column of Table \ref{BNDATA} 
where SC $=1$ for reliable data and SC $=0$ otherwise.
Adding the set of twenty-two selected sources to the sample studied by Jones does not change 
the slope substantially. This can be seen in the Figure 
where the least squares-fit shown by the dashed line still has a slope of $\approx$ 0.75, 
corresponding to the relation
\begin{equation}
P_{K}(\%) = 2.30 \tau_{K}^{0.73}.
\end{equation}
In other words the behavior of the polarization efficiency in OMC-1 is close to the statistical 
behavior observed toward various other LOSs in the Galaxy and the polarization 
efficiency, $P_{K}/\tau_{K} \propto \tau_{K}^{-1/4}$. 

\subsubsection{IR Polarization Map and Histograms of Polarization Data}

\begin{figure}
\epsscale{0.8}
\plotone{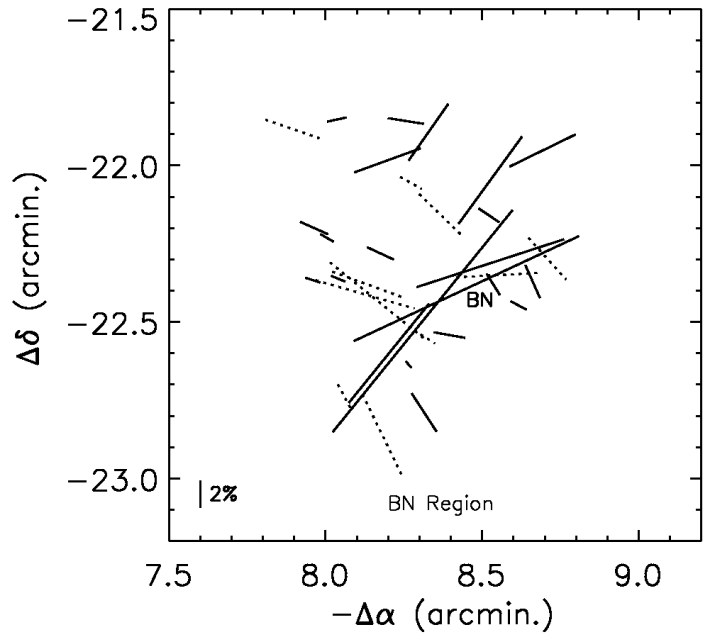}
\caption{$K$-band polarization map in the region of the BN object.
Data are from Table \ref{BNDATA}. 
Solid vectors show reliable data probing the fields while other data are shown with 
dashed vectors. 
The reference position is R.A.=5$^{\rm h}$35$^{\rm m}$48.0$^{\rm s}$, decl.=-5$^{\circ}$ 
00$^{\rm mn}$ 00.0$^{\rm s}$ (J2000.0). 
\label{cartebn}} 

\end{figure}

\begin{figure}
\epsscale{0.8}
\plotone{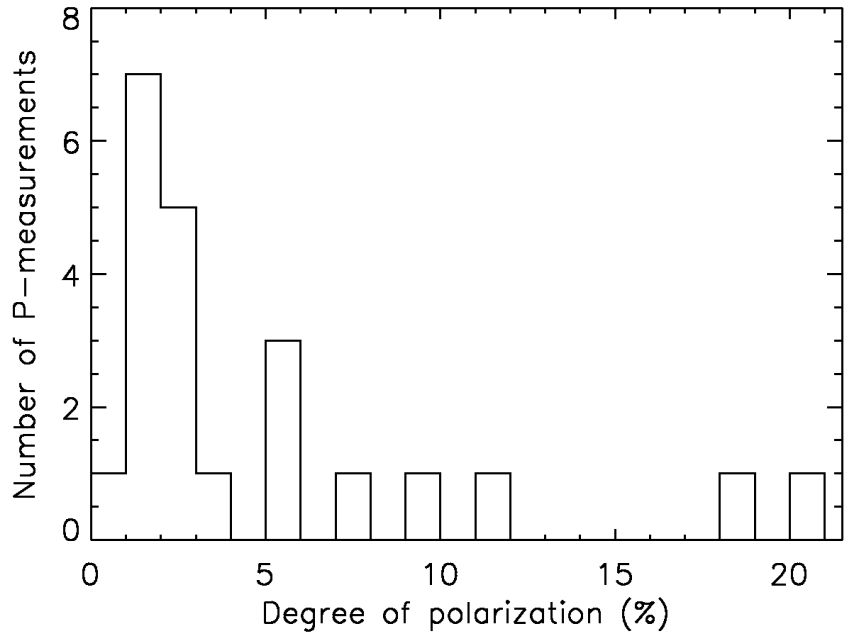}
\caption{Histogram of the $K$-band degrees of polarization in the BN region. 
See Figure \ref{cartebn} and Table \ref{BNDATA}. 
Only reliable data are shown.
Five stars show a polarization degree $> 6 \%$. 
\label{pbn}}
\end{figure}

\begin{figure}
\epsscale{0.8}
\plotone{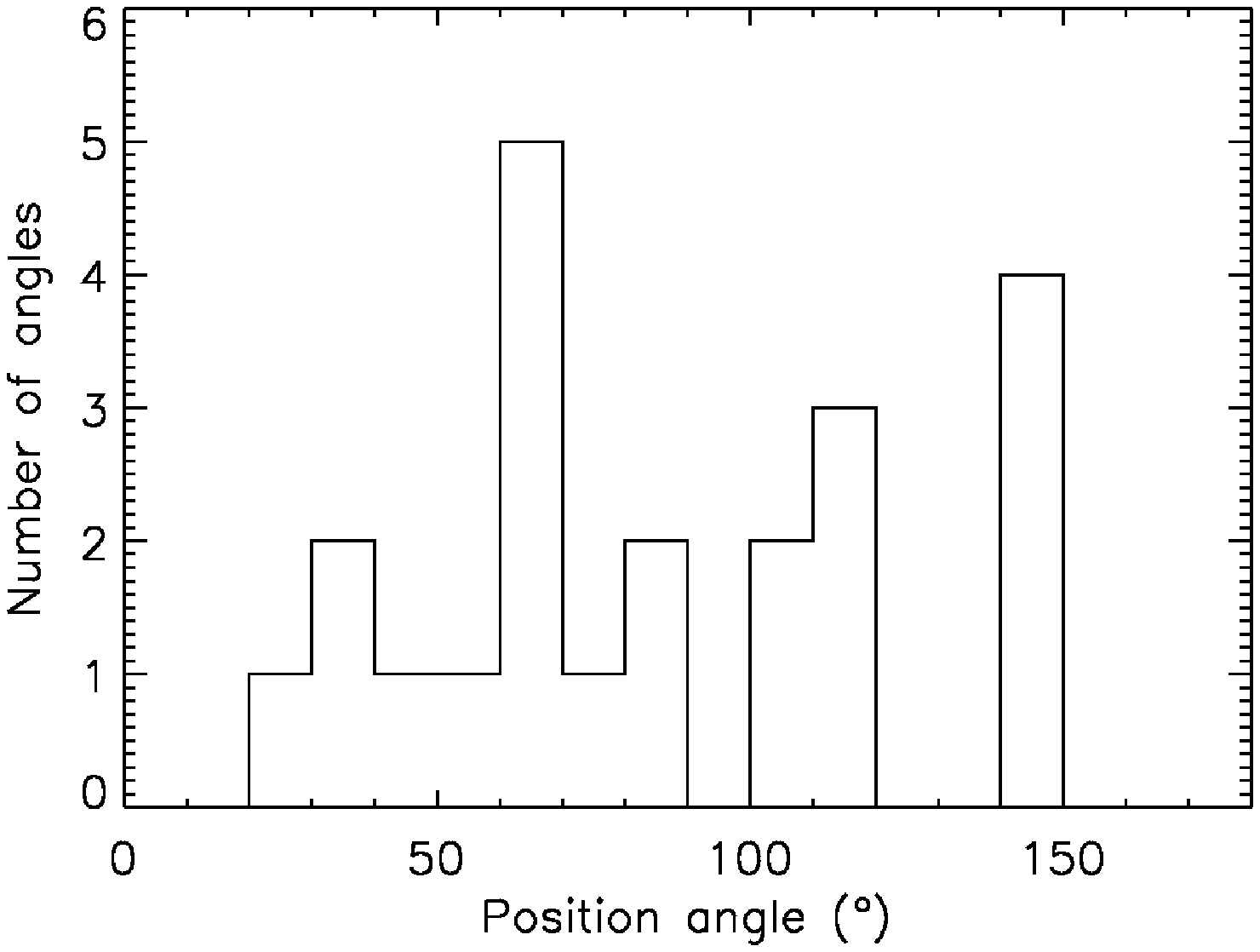}
\caption{
Same as in Figure \ref{pbn} but for position angles.
\label{angbn}}
\end{figure}

All the data from Table \ref{BNDATA} are shown in Figure \ref{cartebn}. 
Data believed to be reliable to probe the magnetic fields are shown with solid
polarization vectors while the other ones are shown with dashed polarization vectors. 
Histograms of polarization percentages and of polarization position angles 
of the reliable sample are shown in Figures \ref{pbn} and \ref{angbn}, respectively.
In Figure \ref{pbn}, the mean and the dispersion of the whole data set (N=22)
are $\overline{P} = 4.89\% $ and $S_{\rm P} = 5.47 \%$, respectively. 
In Figure \ref{angbn}, the mean and the dispersion of the whole data set
are $\overline{\theta} = 84\fdg6$ and $S_{\theta} = 38\fdg5$, respectively.

\section{MULTI-SCALE ANALYSIS OF MAGNETIC FIELDS} \label{comparisons}
      
In this section we compare magnetic fields probed at several wavelengths and on various 
spatial scales. All our basic statistical results are compiled in Tables \ref{tabcomp}
and \ref{tabcomp2}. Each line shows means and dispersions for several regions.
Since the foreground interstellar polarization in this region was estimated by \citet{bas79} to 
be $P_{\rm IS} = 0.53 \pm 0.15 \%$ with $\theta_{\rm IS} = 49^{\circ} \pm 8^{\circ}$, means and dispersions
were systematically estimated for subsets of data with $P \ge 1\%$ and $P<1\%$. In the following, 
except where explicitly mentionned, no foreground correction was applied to the data. 
A rotation of $\pm 90^{\circ}$ was applied to submm data to allow 
direct comparison with visible and NIR polarization position angles. 
All the 850 $\mu$m data have a signal-to-noise ratio such that $P/\sigma_{P}>3$ \citep[][]{poi206}.
All the 350 $\mu$m data have a signal-to-noise ratio such that $P/\sigma_{P} >2$ \citep[][]{hou04}. 

\subsection{Comparison of Results on Large Spatial Scales in Orion A}

\begin{figure}
\epsscale{1.0}
\plotone{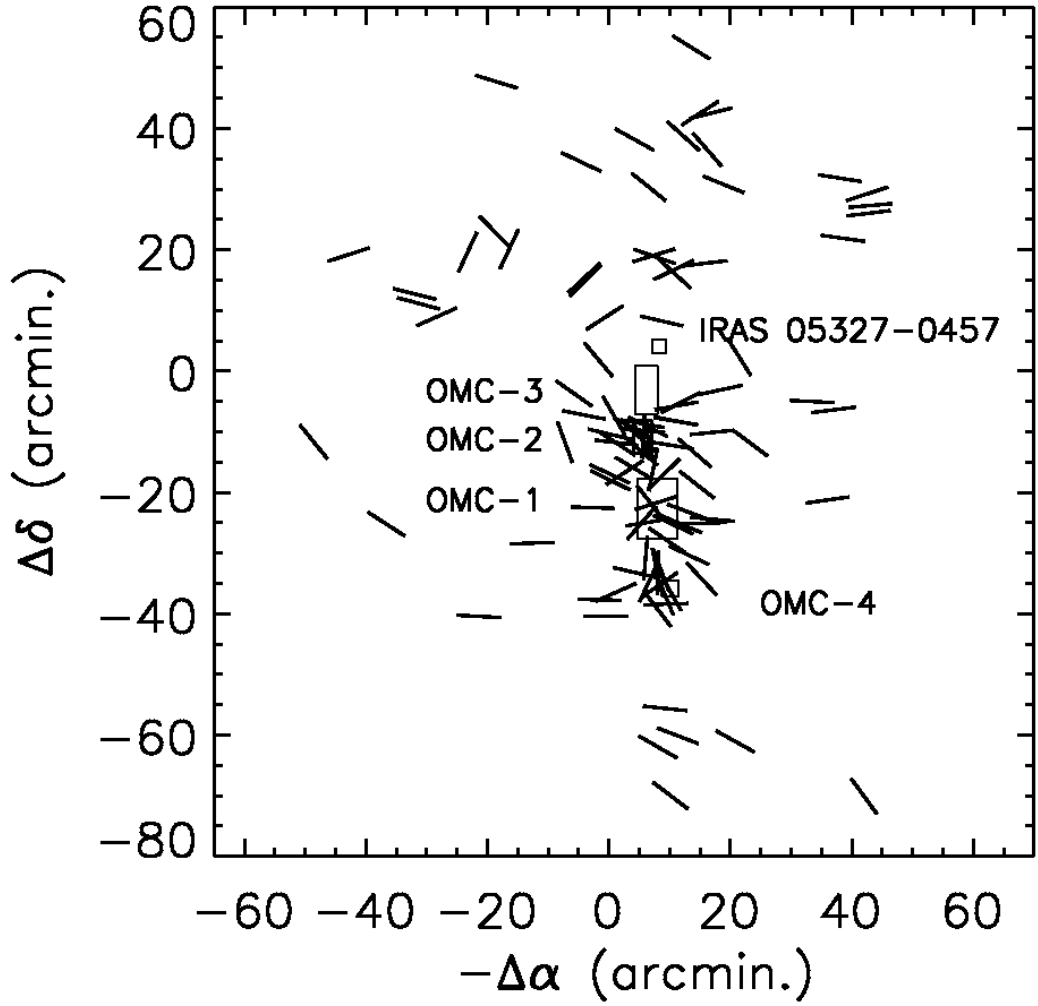}
\caption{Same as in Figure \ref{cartevis} but now all vectors have the same length
for better visualization of the polarization position angles. 
\label{lsunp}}
\end{figure}

All selected data from Tables \ref{OMMDATA}, \ref{PVISDATA}, \ref{OTHERDATA} and \ref{OMC2DATA} are
shown in Figure \ref{lsunp} with all vectors drawn with the same length to get a clear picture 
of visible polarization position angles in Orion A. 
For this region, covering an area of $\approx 1\fdg5 \times 2^{\circ}$, we find a
mean position angle of $\approx 89^{\circ}$ for the more highly polarized stars and a mean position
angle of $\approx 67^{\circ}$ for those with a polarization $P \leq 1\%$ as displayed in column 3 of 
Table \ref{tabcomp}. See also histograms in Figures \ref{anghisto} and \ref{anghisto2}.
Thus we confirm with a larger sample the results found by \citet{bre76}. He found a 
mean position angle of $\approx 100^{\circ}$ with high polarization stars, and a mean 
position angle of $\approx 75^{\circ}$ for low polarization stars, respectively. 
We point out that the mean position angle of $\approx 67^{\circ}$ obtained for stars
with $P \leq 1\%$ is very close to the mean value of about $69^{\circ}$ defined by \citet{li09} 
on an area of about $4^{\circ} \times 5^{\circ}$ (see their Figure 1) by using the catalog 
from \citet{hei00}. Once the polarization results of all our data set
have been corrected for the ISM polarization \citep{bas79} we find a mean polarization 
angle of $\approx 104^{\circ}$ with a dispersion of $\approx 31^{\circ}$. When the same 
is done only on the stars with $P \ge 1\%$, we find a mean polarization 
angle of $\approx 99^{\circ}$ with a dispersion of $\approx 30^{\circ}$.

\subsection{Comparisons on Various Scales Toward OMC-1} \label{comp1}

\subsubsection{Comparison of Submm Data with Visible Data around OMC-1} \label{aroundcomp1}

Figure \ref{compomc1} shows visible data superimposed on the 350 $\mu$m data observed by \citet{hou04}.
All stars labeled in the Figure have reliable data according to our selection criteria $SC=1$ or $2$. 
By referring to Tables \ref{PVISDATA} and \ref{OTHERDATA}, we see that 
polarization is produced by ICD grains and/or that these stars are located at distances 
greater than $500$ pc from the Sun. 
By comparing the corresponding means and dispersions of the data set 
shown in Figure \ref{compomc1} and compiled in column 4 of Table \ref{tabcomp},
we see that apparently no spatial alignment on the POS exists between magnetic fields 
probed with both techniques. The mean visible polarization position angle is $\approx 
64^{\circ}$ with a dispersion of $\approx 39^{\circ}$ while once rotated by $90^{\circ}$ 
the mean submm polarization position angle is at $\approx 112^{\circ}$ with a dispersion 
of $\approx 30^{\circ}$, making a mean offset of about $48^{\circ}$ between the two 
mean uniform field components. We point out that these results are consistent 
with those found by \citet{li09} when visible polarization data are used on a larger scale. 

\subsubsection{Comparison of Submm Data with $K$-Band Data on and around BN} \label{aroundbn}

Figure \ref{compbn} shows our $\lambda = 2.2$ $\mu$m data superimposed on the 350 $\mu$m data 
observed by \citet{hou04}. Here again, by comparing the corresponding means and dispersions 
of the data set shown in the figure and compiled in column 5 of Table \ref{tabcomp},
we find that apparently no spatial alignment on the POS exists between magnetic fields 
probed with both techniques. The mean $K$-band polarization position angle is $\approx 
85^{\circ}$ with a dispersion of $\approx 39^{\circ}$ while once rotated by $90^{\circ}$ 
the mean submm polarization position angle is at $\approx 120^{\circ}$ with a small dispersion 
of $\approx 9^{\circ}$, making a mean offset of about $35^{\circ}$ between the two 
mean uniform field components. We point out, however, that a few stars, among which BN, show a 
polarization position angle consistent with the $90^{\circ}$ rotated submm polarization angles 
in their neighborood.    

\subsubsection{Comparison of  $H$-Band Data with $350$ $\mu$m and $K$-band Data} \label{aroundbnomc1}

Wide-field ($\approx 8^{\arcmin} \times 8^{\arcmin}$) $JHK_{s}$-band polarimetry toward OMC-1 
has been conducted by \citet{kus08}. Their Table 2 shows that the mean position angles
obtained through each filter are consistent with the $90^{\circ}$ rotated 350 $\mu$m 
mean polarization angles of \citet{hou04}.
We used the publicly available $H$-band data of \citet{kus08} and find that apparently none 
of the stars observed at this wavelength matches the stars we observed in the $K$-band.
The mean polarization position angle of the $H$-band data lying in the region shown in 
Figure \ref{compbn} is $\approx 134^{\circ}$ with a dispersion of $\approx 27^{\circ}$.

\subsubsection{A Rotating Magnetic Field on large scale toward OMC-1}

The comparison of the means and dispersions discussed in the preceeding subsections
suggests that the geometry of the magnetic field toward OMC-1 could be relatively complex. 
This result is not a complete surprise since it has been known for a long time that a rotation occurs 
toward the LOS to BN and OMC-1/IRS2, two sources close to each other on the POS \citep[][]{lon80}. 
Recent works, however, give clues to a 
possible rotation on a larger scale around BN and also toward OMC-1. 
Circular Polarization (CP) was obtained in the $K_{s}$-band by \citet{fuk09} on a region 
of $2\farcm7 \times 2\farcm7$ centered close to the position of BN. 
The authors find a good correlation between CP and extinction by using $H-K_{s}$ as a proxy
for extinction. Their analysis also shows a good correlation between CP and linear polarization,
and dichroic extinction would be the main mechanism producing the CP observed in this region.
In addition, because of the well defined uniform submm polarization
pattern, it seems reasonable to assume that turbulent effects can be neglected. This hypothesis is 
supported by the results of the angular dispersion analysis of the 350 $\mu$m polarization 
data of OMC-1 by \citet{hil09} where the uniform magnetic field component on the POS 
is estimated to be ten times stronger than the turbulent magnetic field component in high density 
regions of the clouds. 

In order to check the hypothesis of a rotation of the magnetic fields on a larger scale around BN 
and toward OMC-1 we have measured the offset in position angle between 
each $K$-band polarization vector and the 4 nearest submm polarization vectors located in their neighborhood 
(see Figure \ref{compbn}). 
Present grain alignment theory suggests that submm polarization vectors are perpendicular to the 
mean magnetic field direction on the POS, while visible polarization vectors should be parallel 
to this mean direction, so one would expect a spatial correlation between 
some components of the magnetic fields probed with both techniques. 
The histogram of the angular offset estimates is shown in Figure \ref{histocompbn}. 
We are concerned only with the magnitude of the difference between the visible and submm
polarization vectors, therefore we convert all offsets to lie between $0^{\circ}$ and $90^{\circ}$.   
If submm polarimetry probes mainly magnetic fields embedded into the clouds, 
a similar orientation of the fields probed at each wavelength would correspond to an angular offset 
of $90^{\circ}$. We find a peak around $70^{\circ}$ suggesting that some 
stars observed in the $K$-band probe the same magnetic field component than submm data do. 
Other stars, however, show offsets between $0^{\circ}$ and $50^{\circ}$
implying that other magnetic field directions exist along LOSs toward OMC-1. 

It was shown that NIR polarimetry at BN is dominated by dichroism  
and that the $\lambda = 2.2$ $\mu$m degree of polarization and position angle are consistent 
with $100$ $\mu$m data \citep[see][]{lon80,lee85,jon89,hou96,sch98,jia05}.
In addition, in our sample BN has the second highest $K$-band polarization degree 
and the highest extinction parameter, therefore one would expect to see 
an increase of the degree of polarization, $P_{K}$, as well as an increase of the optical depth, 
$\tau_{K}$, for angular offsets closer and closer to $90^{\circ}$.
This result is suggested in Figure \ref{ptauoffbn} with 
our relatively small sample of 22 measurements. 
The top part of Figure \ref{ptauoffbn} shows values of $P_{K}$ up to $\approx 20\%$ for angular 
offsets lying between $\approx 65^{\circ}$ 
and $\approx 90^{\circ}$ whereas values of $P_{K} \lesssim 5\%$
are found for angular offsets outside this range. 
The same trend is suggested in the bottom part of Figure \ref{ptauoffbn}
where the highest values of $\tau_{K}$ are found for angular offsets around $90^{\circ}$
whereas most of the values of  $\tau_{K}$ lying between $\approx 0^{\circ}$ and 
$\approx 65^{\circ}$ are two to three times lower than those values.

As a further test of this idea, we used the larger $H$-band data sample of \citet{kus08}.
Here again, we have measured the offset in position angle between 
each IR polarization vector and the 4 nearest submm polarization vectors located in their neighborhood. 
Because the 350 $\mu$m data are gridded with 20$\arcsec$ pixels we have restricted our calculations 
to distances smaller than 21$\arcsec$ on the POS.
This ensures that we avoid $H$-band data lying too far on the POS from the area mapped at 350 $\mu$m. 
The histogram of the angular offsets (top) and the distribution of $P_{H}$ (middle) with those angular offsets 
are shown in Figure \ref{hoffsets}. An increase 
of the polarization degree, $P_{H}$, as a function of the angular offset can be seen in this figure.   
This result confirms the trend observed around BN with the smaller $K$-band data sample 
(Figure \ref{histocompbn}) and implies that the magnetic field could rotate from inside-out 
toward OMC-1 on a spatial scale larger than just the BN area  
of about $1\farcm5 \times 1\farcm5$.  
As a statistical check, we have calculated the mean polarization degrees in bins of $18^{\circ}$.
The bottom part of Figure \ref{hoffsets} shows a smooth increase of $<P_{H}>$ from $2.5\%$ to 
$7.0\%$ for offsets increasing from $\approx 50^{\circ}$ to $\approx 80^{\circ}$ 
Within the uncertainties on the polarization position angles 
this trend is consistent with a rotation of the magnetic field of $\lesssim 60^{\circ}$ 
on large scale toward OMC-1.
For angular offsets between 0$^{\circ}$ and 50$^{\circ}$ $<P_{H}>$ 
is approximately constant at $\approx 2.5\%$ as illustrated by the horizontal dashed line.

\subsection{Comparisons of Magnetic Fields in and around OMC-2 and OMC-3}
 
The visible polarization data selected in Tables \ref{OMMDATA} to \ref{OMC2DATA} are shown in Figures 
\ref{compomc23} and \ref{compomc23zoom}. They are superimposed on the 850 $\mu$m polarization data 
observed and discussed by \citet{poi206}. In the two maps all vectors are displayed with a similar 
size for a better vizualization of the polarization position angles but two different size scales 
are used. Bold vectors show visible data and thin vectors show submm data. 

Figure \ref{compomc23} shows the spatial distribution of the visible polarization vectors 
in and around OMC-2 and OMC-3 with respect to the polarization pattern produced by dust emission within the clouds. 
The means and dispersions relative to the two data sets shown in the Figure are 
given in column 3 in Table \ref{tabcomp2}. Statistical values of the 350 $\mu$m polarization 
data discussed by \citet{hou04} are also displayed in the Table. The two submm data sets 
show different means and dispersions. The mean orientation at 350 $\mu$m is
$\approx 45^{\circ}$ with a dispersion of $\approx 25^{\circ}$ while
the 850 $\mu$m data set has a mean of $\approx 56^{\circ}$ with a dispersion 
of $\approx 36^{\circ}$. Part of this difference is probably due to the smaller number 
of 350 $\mu$m measurements compared to the 850 $\mu$m data set and to some differences
in the regions covered by each map. For a detailed comparison between the two 
submm data sets see \citet{poi206}.
When looking from visible to submm data, the results 
show an anticlockwise rotation of $\approx 11^{\circ}$ and $\approx 22^{\circ}$
in the mean position angle when 850 $\mu$m and 350 $\mu$m data are considered, respectively.
These results show good agreement between the mean orientation of the magnetic fields 
probed with both techniques in and around OMC-2/OMC-3. 

Figure \ref{compomc23zoom} shows a zoom on the two data sets displayed toward OMC-2.
The means and dispersions for each subset of data shown in the Figure are compiled in column 4 in Table 
\ref{tabcomp2}. The $90^{\circ}$-rotated mean polarization position angle of the 350 $\mu$m pattern is 
found to be similar to the mean position angle of the visible data, $\approx 48^{\circ}$. 
However, the set of 850 $\mu$m data gives a different mean position angle of $\approx 69^{\circ}$ 
with a large dispersion. These results suggest that magnetic fields probed with both techniques 
have a similar orientation projected on the POS, taking the small number of visible 
measurements ($N$ = 9) into account.

Our results are generally consistent with those for OMC-2 and OMC-3 shown by \citet{li09} in their Table 1.

\subsection{Comparisons of Magnetic Fields in and around OMC-4}

Our results for this region are compiled in column 5 in Table \ref{tabcomp2}.
The mean polarization position angle of visible data covering the area shown 
in Figure \ref{compomc4} is $\approx 54^{\circ}$ with a dispersion 
$S_{\theta}$ $\approx 36^{\circ}$. The $90^{\circ}$-rotated 
mean polarization position angle of the well defined 350 $\mu$m polarization pattern is  
$\approx 76^{\circ}$ which yields a difference of $\approx 22^{\circ}$ between 
the magnetic field probed with visible and submm techniques on small scale around 
OMC-4. These values are consistent with the position angle of $69^{\circ}$ for the 
ISM magnetic field on a larger scale inferred by \citet{li09}.

\section{DISCUSSION} \label{discussion}

The main trend of the comparative studies presented in section \ref{comparisons} is that 
magnetic fields probed with submm techniques inside the molecular star-forming ISF
have directions generally consistent with those
probed with visible techniques in the diffuse parts of the surrounding ISM, except
for OMC-1 as discussed above in section \ref{comparisons}.

\begin{figure}
\epsscale{1.0}
\plotone{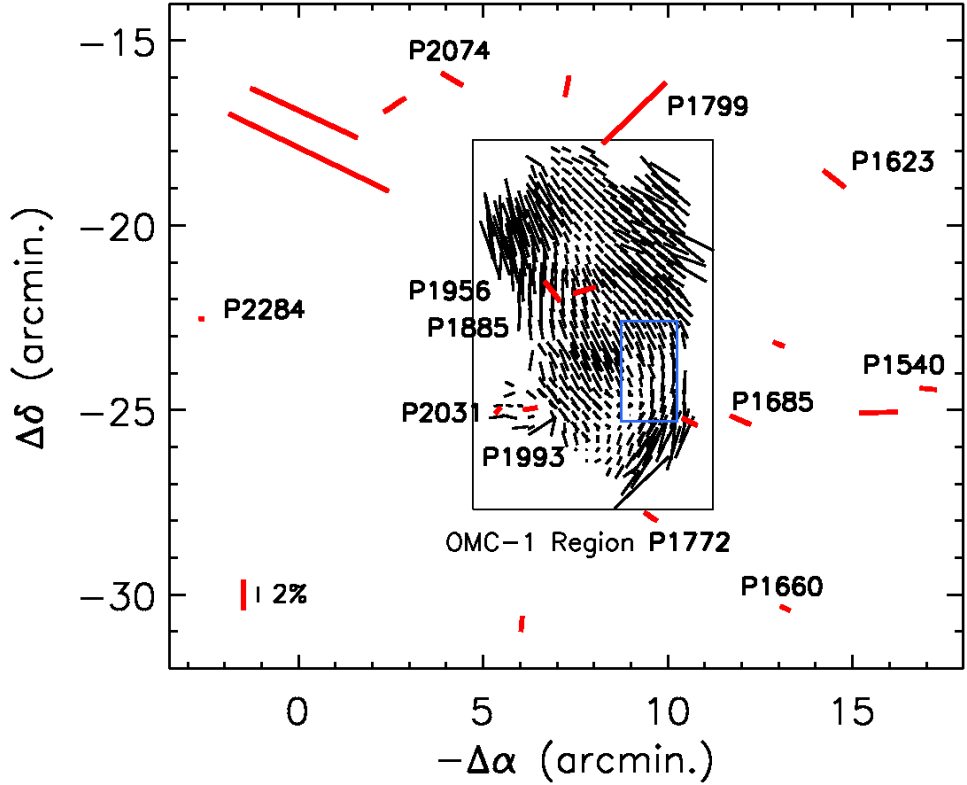}
\caption{Visible and $350$ $\mu$m polarization map of OMC-1. The black rectangle delineates OMC-1. Submm data 
traced with thin vectors are from \citet{hou04}. The reliability of the data refers to their
reliability to probe magnetic fields (see section \ref{pvisclass} and \ref{pvisext} for details).
Named stars have reliable data according to this definition. 
The red vector located close to the middle of the submm map refers to star P1885.
The blue rectangle shows the SW region observed by \citet{bus05} (see discussion for details).
The reference position is R.A.=5$^{\rm h}$35$^{\rm m}$48.0$^{\rm s}$, decl.=-5$^{\circ}$ 
00$^{\rm mn}$ 00.0$^{\rm s}$ (J2000.0). 
\label{compomc1}}
\end{figure}

For OMC-1, \citet{sch98} compared the field lines probed with visible polarimetry 
data from \citet{bre76} and \citet{vrb88} in the ambient envelope directly outside the OMC-1 ridge
with the field lines probed with 100 $\mu$m and 350 $\mu$m polarimetry data inside the clouds.
He showed some similarity between the orientation of some components probed at each wavelength 
(as one can see also in our Figure \ref{compomc1}). In addition the magnetic field component
probed with IR polarimetry in the LOS to BN was found to be consistent with the one probed with 
submm data in the same LOS. 

Our analysis presented in section \ref{comp1} fills the gap between the two 
scales discussed by \citet{sch98} and suggests that a large scale rotating magnetic 
field could be wrapping the OMC-1 clouds. The results 
suggest a rotation by $\lesssim 60^{\circ}$ toward OMC-1 (Figure \ref{hoffsets}).
This is the first time that this trend has been shown with NIR data over such a wide field.
This is also the first time that this twist has been shown to extend so deep into the cloud
with linear polarimetry. 

Based on a previous analysis by \citet{mar74} and on large circular polarization measurements at 
$\lambda = 2.2$ $\mu$m combined to linear polarimetry, \citet{lon80} argued 
that the polarization of OMC-1/IRS1 (BN) and OMC-1/IRS2 
is produced by a medium of aligned grains with a twist in the alignment 
$ \lesssim 60^{\circ}$, providing support on a small scale to our analysis.
More recent works by \citet{hou96} and by \citet{ait06} 
also support this scenario along the LOS to BN.
On a larger scale, based on the $K$-band and $H$-band coupled to submm data analysis shown in 
Figures \ref{ptauoffbn} and \ref{hoffsets}, we think that the trend 
observed in those figures shows a rotation of the polarization plane toward OMC-1. 
Whether this rotation is produced by a two- or multi-slab model or by a spatially continuous 
rotation toward OMC-1 is not clear. Such possibilities were 
discussed by \citet{ait06} and references therein. 
We now look at the 350 $\mu$m polarization pattern observed by \citet{hou04} 
and into the literature to find out if one scenario could prevail. 

\begin{figure}
\epsscale{0.8}
\plotone{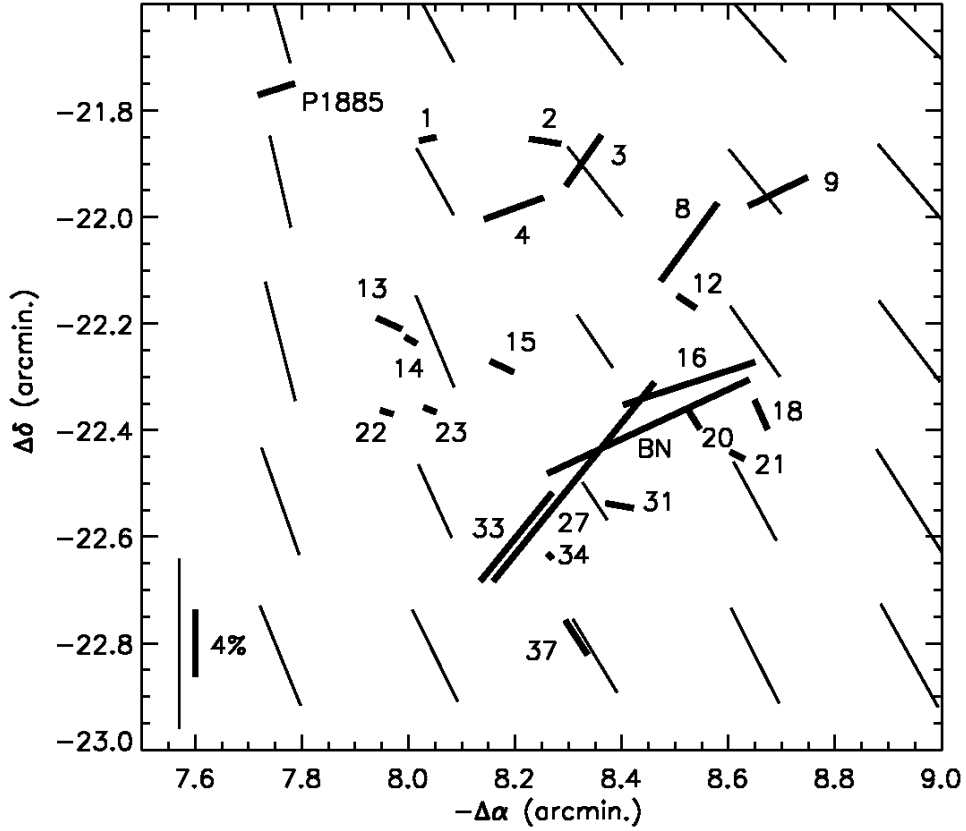}
\caption{IR ($\lambda = 2.2$ $\mu$m) and submm ($\lambda = 350$ $\mu$m) polarization map 
through the BN region. Submm data are from \citet{hou04}. 
Selected $K$-band data from Table \ref{BNDATA} are shown with bold lines.
Also shown in the map is the visible measurement on P1885 from Table \ref{PVISDATA}. 
Submm data are shown with thin lines. Both scales are given at the 
bottom left in the Figure. This map can be compared with the one shown in 
Figure \ref{cartebn}. BN is located at 
($\alpha(2000),\delta(2000)$)=(5:35:14.2,-5:22:23.6)=($-8.45^{\arcmin}$,$-22.39^{\arcmin}$).
The reference position is R.A.=5$^{\rm h}$35$^{\rm m}$48.0$^{\rm s}$, 
decl.=-5$^{\circ}$00$^{\rm mn}$ 00.0$^{\rm s}$ (J2000.0). 
\label{compbn}}
\end{figure}

\begin{figure}
\epsscale{0.8}
\plotone{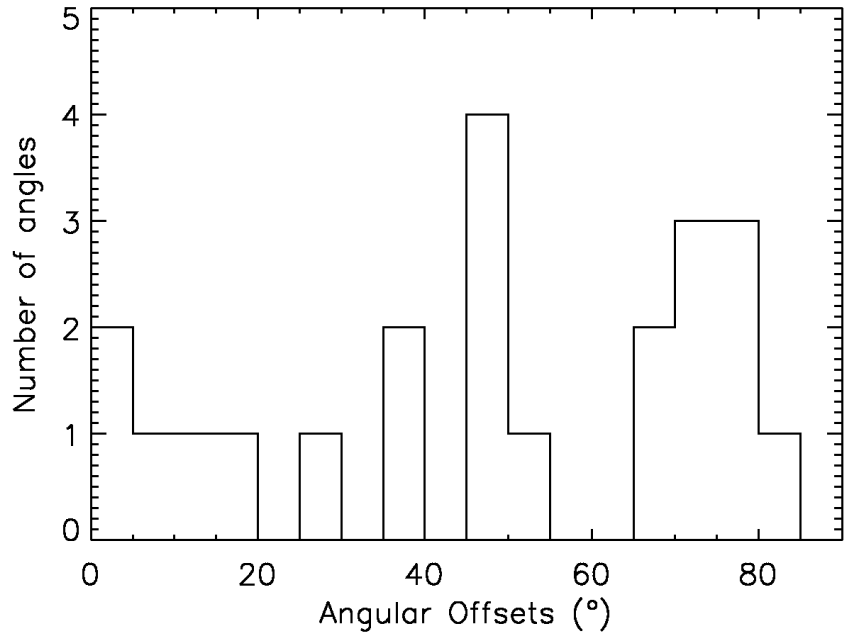}
\caption{Histogram of the mean offsets of polarization position angles between 2.2 $\mu$m and 350 $\mu$m 
data shown in Figure \ref{compbn}. 
\label{histocompbn}}
\end{figure}

\begin{figure}
\epsscale{0.8}
\plotone{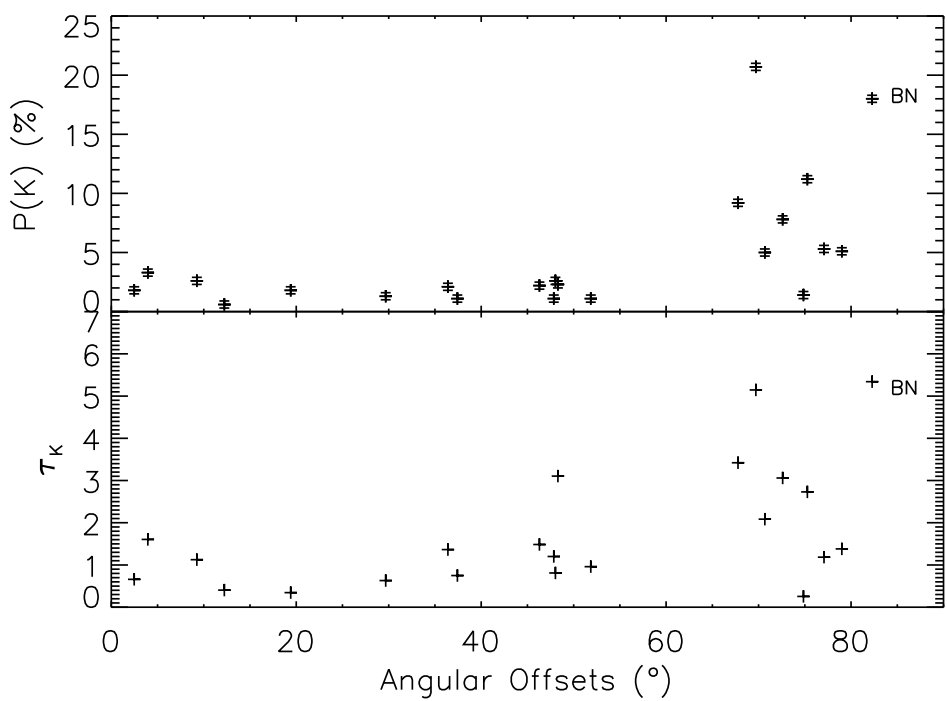}
\caption{$P_{K}$ versus $\theta_{\rm off}$ (top)
and $\tau_{K}$ versus $\theta_{\rm off}$ (bottom)
for sources selected in Table \ref{BNDATA} and shown in Figure \ref{compbn}. 
We find an increase of both the degree of polarization, $P_{K}$, and the optical depth, 
$\tau_{K}$, for angular offsets around $90^{\circ}$, indicative of a rotation
of the magnetic field toward LOSs to BN and its vicinity.
\label{ptauoffbn}}
\end{figure}

\begin{figure}
\epsscale{0.8}
\plotone{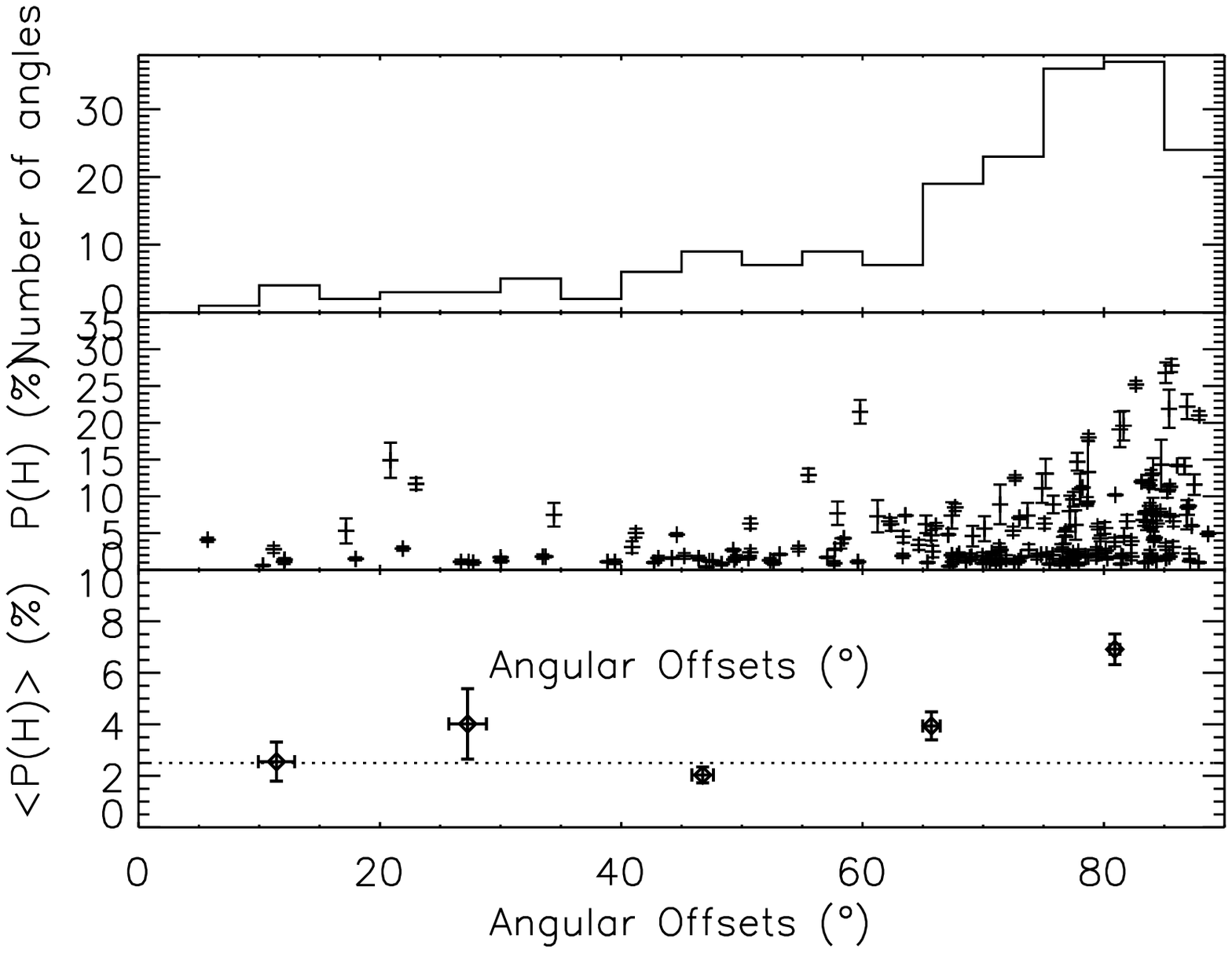}
\caption{Histogram of the angular offsets between $H$-band polarization data from \citet{kus08} 
and 350 $\mu$m data from \citet{hou04} (top),
the distribution of $P_{H}$ with these angular offsets (middle) and 
the distribution of $<P_{H}>$ with mean angular offsets in bins of 18$^{\circ}$ (bottom).
All selected $H$-band data have a polarization percentage and uncertainty such that $P/\sigma_{P} > 3$. 
$<P_{H}>$ increases as a function of the angular offsets between $\approx 50^{\circ}$ and $\approx 90^{\circ}$, 
and $<P_{H}> \approx 2.5\%$ outside this range (horizonal dashed line).
\label{hoffsets}}
\end{figure}

\begin{figure}
\epsscale{1.0}
\plotone{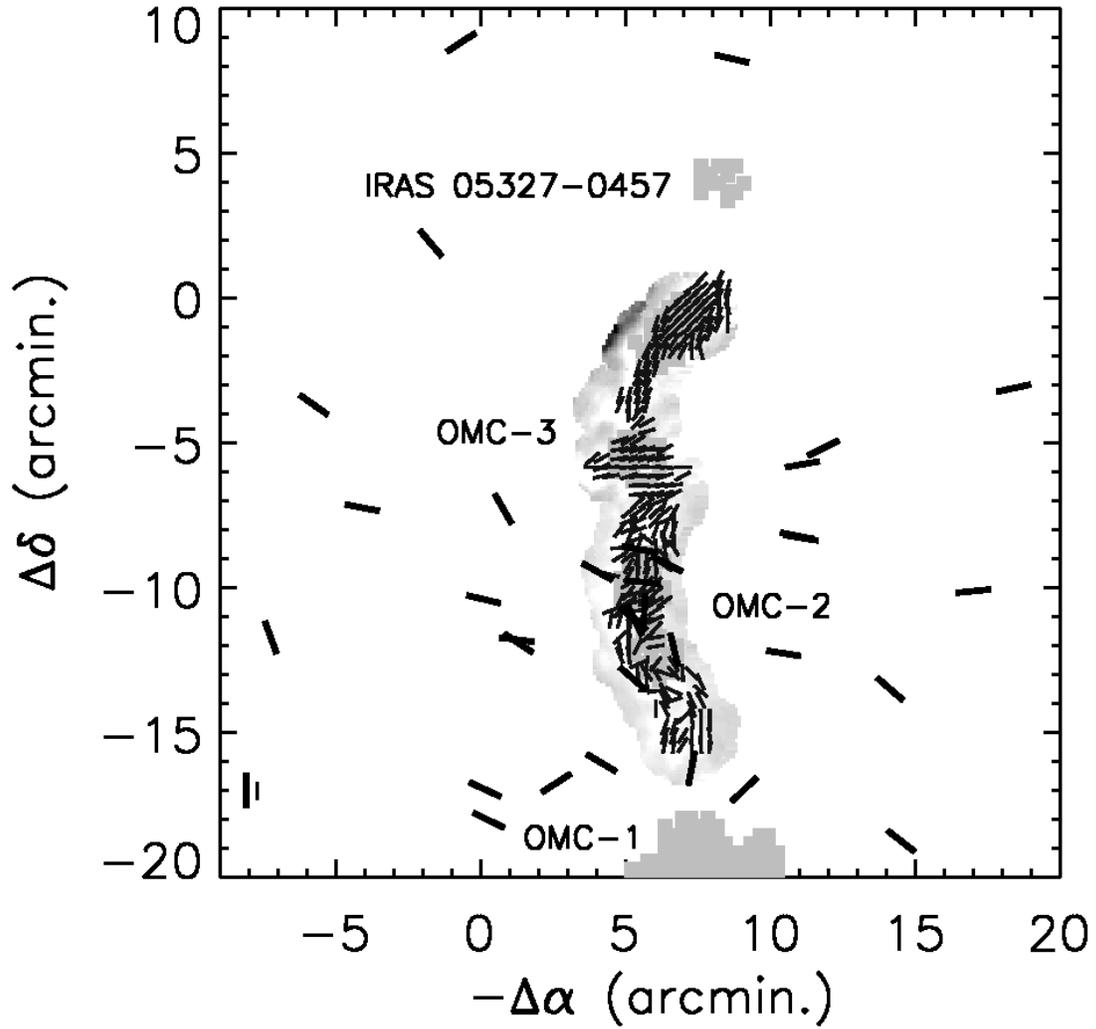}
\caption{A map of 850 $\mu$m data \citep{poi206}
superimposed on visible data. 
Vectors in each wavelength regime have the same length for better visualization 
of the polarization position angles, with the scales given in the bottom left.
Bold vectors: visible data. Thin vectors: submm data. The reference position is 
R.A.=5$^{\rm h}$35$^{\rm m}$48.0$^{\rm s}$, 
decl.=-5$^{\circ}$00$^{\rm mn}$ 00.0$^{\rm s}$ (J2000.0). 
\label{compomc23}}
\end{figure}

\begin{figure}
\epsscale{0.8}
\plotone{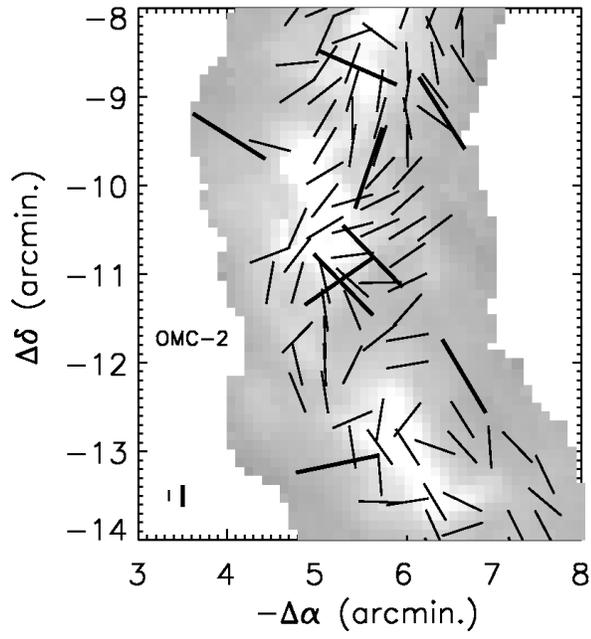}
\caption{Same as in figure \ref{compomc23} but with a zoom on regions FIR3 to FIR6. 
\label{compomc23zoom}}
\end{figure}

\begin{figure}
\epsscale{0.8}
\plotone{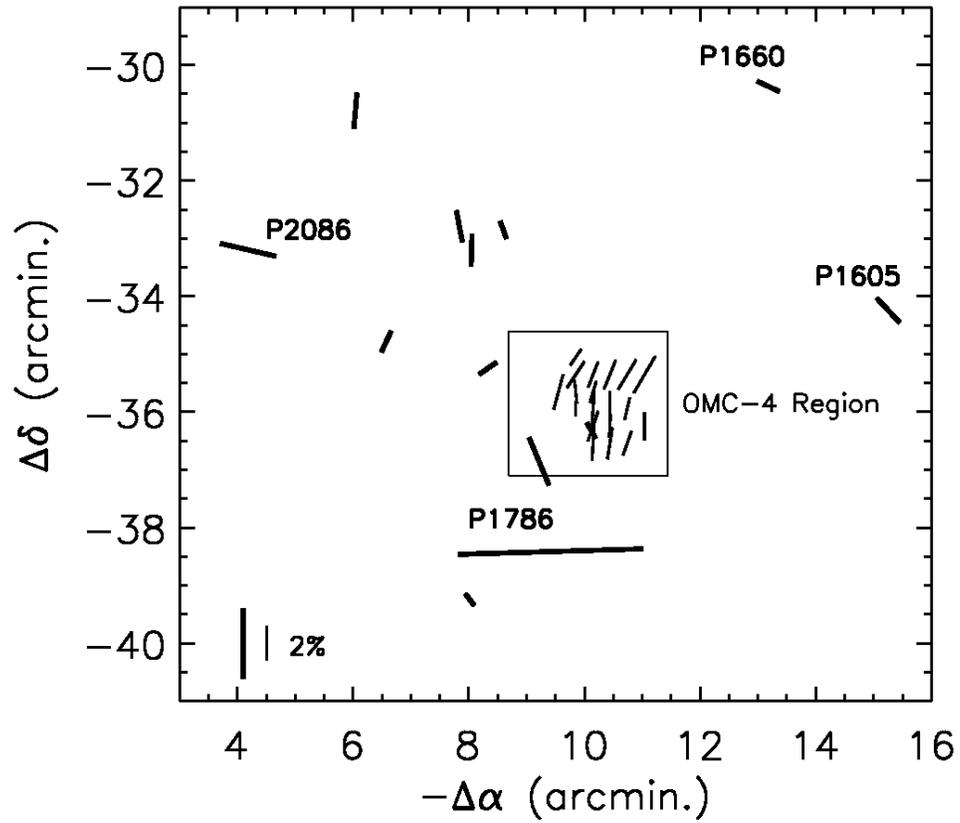}
\caption{Map of 350 $\mu$m data from \citet{hou04} (thin vectors) superimposed on visible data (bold vectors). 
Polarization scales are given in the bottom left.
The reference position is R.A.=5$^{\rm h}$35$^{\rm m}$48.0$^{\rm s}$, 
decl.=-5$^{\circ}$00$^{\rm mn}$ 00.0$^{\rm s}$ (J2000.0). 
\label{compomc4}}
\end{figure}

We divide the 350 $\mu$m polarization map into two main regions showing smooth 
polarization patterns with different mean orientations. 
The first region is located in the 
south-western part of the map where $\delta \la -5^{\circ} 24{\arcmin}$ (see Fig. \ref{compomc1}). 
Most if not all the 350 $\mu$m data in this region show a polarization position angle $\theta > 90^{\circ}$. 
This distribution yields a mean magnetic field direction of $\overline{\theta_{\rm B}} \approx 60^{\circ}$.
Interestingly, this mean orientation is close to the one found with 
visible data in the vicinity of OMC-1 (see column 4 of Table \ref{tabcomp})
along LOSs of small visual extinction. For this reason, 
in the following we call this region of the cloud the near region.
Going north-east of this region, the polarization falls to zero, 
and then the position angles turn by an angle $ \approx 90^{\circ}$. 
Beyond the strip of (almost) null vectors, 
mainly in the northern part of the map, 
most of the 350 $\mu$m vectors have a polarization position angle such that $\theta < 90^{\circ}$.
Those data yield a mean magnetic field orientation $\overline{\theta_{\rm B}} \approx 123^{\circ}$. 
This component would be mostly background to the cloud, 
in agreement with the NIR position angle observed on BN (see Fig. \ref{compbn}).
In the following we call this region the far region.

In Figure 4 of \citet{kus08}, where $\delta > -05^{\circ}23{\arcmin}$, most of their blue vectors match well 
with the far region, some of them showing high polarization values. South of this declination, 
the IR polarization vectors matching well the far region are those showing a `small' polarization degree 
of 1-3$\%$. Such a low polarization could be explained by superposition of the near region mean field 
component at $\overline{\theta_{\rm B}} \approx 60^{\circ}$ but dominated by the far region field component 
along these LOSs. This decrease of polarization is supported by our angular offsets analysis displayed 
in Figure \ref{hoffsets}. \citet{kus08} suspect multiple field components in this region, known as the bar, and 
due to a photodissociation front. \citet{hou04} found an inclination of 49$^{\circ}$ for the magnetic 
field near the bar with respect to the LOS, which would also contribute to a lower polarization. 

The separation of the submm polarization pattern into two regions
discussed above suggests the existence of 
at least two slabs of aligned dust grains toward OMC-1. If these slabs
extend all over OMC-1 one expects to observe Circular Polarization (CP) produced mainly by 
birefringence as well as Linear Polarization (LP) produced by dichroic absorption \citep[e.g.][]{mar74}. 
The HST NICMOS polarization measurements of OMC-1 at $\lambda = 2.0$ $\mu$m with $0.2''$
resolution by \citet{sim06} and covering $\approx 20'' \times 40''$ on and around BN
show that although polarization produced by scattering is present around several 
sources, a component produced by dichroic extinction appears to dominate 
through several lines of sight. On a slightly larger scale 
\citet{fuk09} searched for a correlation between CP and LP on a region covering  
$\approx 2\farcm5 \times 2\farcm5$ centered around BN. 
Their observations show a good correlation between 
the degree of LP in the $K_{s}$ band and $H - K_{s}$ used as a proxy for extinction. 
Their study also shows a correlation between the degree of CP and the degree of LP
oriented at $\pm 45^{\circ}$ relative to the grain alignment axis but a lack of correlation between 
CP and LP parallel or perpendicular to the grain axis which is indicative that CP is mainly
produced by birefringence. On a larger scale in OMC-1,
\cite{bus05} show that the degree of CP correlates with redenning in
the molecular cloud, as measured by the $J-K$ color index. 
CP appears to be generally very low toward less dense regions dominated 
by H$_{\rm II}$, in agreement with the correlations found by \citet{fuk09}. 

We suspect that the zone where the 350 $\mu$m polarization is almost null corresponds to LOSs where 
the effects of the two large scale components cancel mutually. Such an effect  
where two regions have a different grain alignment along the LOS is expected to 
give rise to circular polarization. This is confirmed by the CP 
map of \citet{bus05} in their figure 3 where one can see knots of high 
CP in various locations along the strip which corresponds to null submm polarization. 
The area covered by their Figure 3 is shown by the blue rectangle in our Figure \ref{compomc1}. 
The end of the strip where the submm polarization is null is located in the lower 
left corner of that blue box. On the contrary, two regions of higher CP showing higher 
submm LP are seen in the upper part of the Figure. One of those was observed 
by \citet{chr00} \citep[see also Figure 1 of][]{bus05}. In this region \citet{mats09} 
show that the high degree of CP could be explained by scattering of highly 
linearly polarized incident light.  

The schematic model that emerges is shown in Fig. \ref{helicalmap}.
The 850 $\mu$m map of \citet{joh99} is used to show the ISF. The symmetry axis of the two large 
scale magnetic field components seen in the $350$ $\mu$m data is suggested by the vertical translucent line. 
The morphology of the large scale magnetic field pervading OMC-1 is shown by the large scale two component 
structure. The foreground component is shown with a full line while the background component 
is shown with a dashed line. The drawn by eye green-blue vectors show the mean orientation of the 
magnetic field probed by submm data in several subregions. 
The black circles indicate regions where the foreground and background magnetic field components 
probably cancel each other within OMC-1.
The crosses (right) and head arrows (left) indicate the direction of the magnetic field with 
respect to the LOS as suggested by Zeeman measurements \citep[see Fig. 15 in][]{hei97}. Crosses 
indicate magnetic fields pointing away from the observer and head arrows magnetic fields pointing toward 
the observer. The rotation that is produced when moving from the far region, i.e. OMC-1, 
toward us on the LOS is illustrated by the blue arrows. The sense of the rotation is based
on the combination of the two following assumptions. The twist of the magnetic field along the LOS to BN 
of $ \lesssim 60^{\circ}$ \citep[][]{mar74,lon80} is indicative of the amount of rotation 
over the whole OMC-1 region. The vectors shown in Figure \ref{compbn} with a polarization 
degree of a few percent (e.g. sources 13, 15, 21) are associated to the near region while 
those showing high polarization degree (e.g. sources 8, 20, BN) are associated to the far region. 
Combining all the preceeding information the large scale magnetic field wrapping OMC-1 should be oriented 
from north to south. It is illustrated by the arrows on the solid and dashed lines. 
Such an orientation would be consistent with a reversal of the field lines
direction toward OMC-1 as suggested by Zeeman measurements, and with the rotation of the magnetic 
field by $\lesssim 60^{\circ}$ toward OMC-1.  

This two-component magnetic field model is not perfect but it does explain a significant fraction of 
our data. Such a large scale magnetic field pattern is compatible with the results of 
\citet{hil09} that uniform magnetic field components dominate over turbulent 
components within OMC-1. Such a large scale structure is also in 
agreement with the statistical analysis of \citet{li09} on 25 molecular clouds which shows a significant 
correlation between the giant molecular clouds accumulation scale and the cloud cores scale. A comparison 
of their results with simulations favors sub-Alfv\'enic cases where magnetic fields play a significant role. 

There is a significant possibility that this two-component model extends to the north beyond the OMC-1 region. 
\citet{mwf01} interpreted their OMC-3 submm data in terms of a helical field. 
\citet{poi206} showed that the uniform magnetic field component is dominant in OMC-3, and 
comparable to the turbulent component in OMC-2. It will be very interesting to find out if these 
results are confirmed by upcoming future observations with POL-2 and SCUBA-2 at the JCMT. 

\section{SUMMARY} \label{sum}

Polarimetry in a broad red bandpass on 69 stars, 41 of which had never been observed 
polarimetrically before, was combined with polarimetry in the $V$, $R$ and $I$ bands on 14 stars
through and around OMC-2, and with $K$-band polarimetry on 33 stars through OMC-1. Together 
with published measurements, these measurements led to a representation of the morphology of 
magnetic fields through and around the filaments and star forming regions OMC-1, OMC-2, OMC-3 
and OMC-4 in Orion A. Our main conclusions are summarized as follows.

1. Magnetic fields probed inside OMC-2, OMC-3 and OMC-4 in the submm are generally consistent 
with those probed in the diffuse parts of the ISM in the surroundings of Orion A with visible polarimetry. 
\citet{li09} conducted the same analysis at scales lying between the accumulation scale length 
and the size of cloud cores.
They found a significant correlation for field directions detected from the two extremes. 
Our analysis confirms their results on intermediate scales.

2. In OMC-1, we measured the angular offsets between the $K$-band 
and the $350$ $\mu$m polarization data. We find high values of the polarization degree, $P_{K}$, and 
the optical depth, $\tau_{K}$, close to angular offsets of $90^{\circ}$ 
whereas lower values of $P_{K}$ and $\tau_{K}$ are observed for smaller offsets.
On a larger scale, we measured the distribution of angular offsets between available $H$-band polarization data 
and the same submm data set. Here we find an increase of $<P_{H}>$ 
with angular offset which we interpret as a rotation of the magnetic field by $\lesssim 60^{\circ}$. 
These trends generalize results that were known only toward the LOSs to BN and OMC-1/IRS2
\citep[e.g.,][]{lon80,hou96,ait06} and are consistent with a twist of the magnetic field on a
larger scale toward OMC-1.

\begin{figure}
\epsscale{0.8}
\plotone{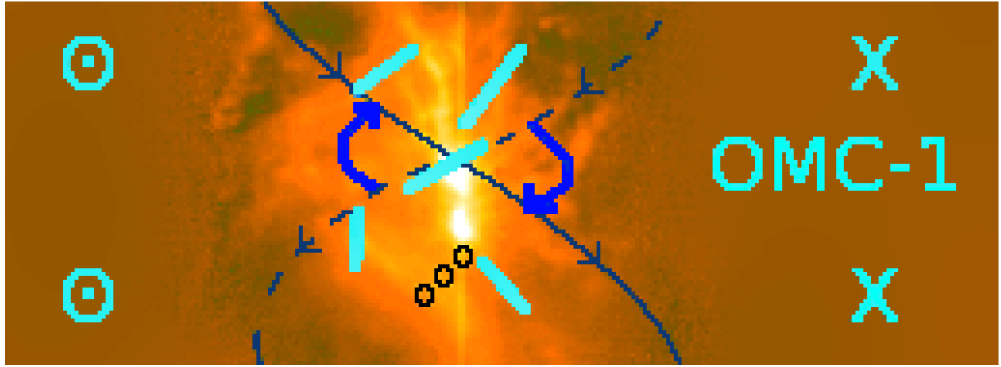}
\caption{A schematic view of the magnetic field that could be wrapping the OMC-1 filament. 
The 850 $\mu$m map of \citet{joh99} is used to show the ISF. The symmetry axis of the two large 
scale magnetic field components seen in the $350$ $\mu$m data is suggested by the vertical translucent line.  
The morphology of the large scale magnetic field pervading OMC-1 is shown by the large scale two component structure.
The foreground component is shown by a full line while the background component is shown with a dashed line.
The drawn by eye green-blue vectors show the mean orientation of the magnetic field probed by submm data in several subregions. 
The black circles indicate regions where the foreground and background magnetic field components 
probably cancel each other within OMC-1.
The rotation that is produced when moving from the far region, i.e. OMC-1, toward us on the LOS 
is illustrated by the blue arrows. 
Such a magnetic field could be blended to the external field probed with Zeeman data \citep{hei97}.
The field comes out of the figure on the east side (arrows) and goes into it on the west side (crosses). 
\label{helicalmap}}
\end{figure}

3. A comparison of our results with several other studies suggests that a two-component magnetic 
field or an helical magnetic field could be wrapping the OMC-1 filament, as illustrated 
in Fig. \ref{helicalmap}. The background component is oriented mostly at  
$\overline{\theta_{\rm B}} \approx 120^{\circ}$ on the POS, and the foreground component is 
oriented mostly at $\overline{\theta_{\rm B}} \approx 60^{\circ}$.
This magnetic field model is compatible with the direction of the field along the LOS determined from 
Zeeman measurements around Orion A \citep{hei97}. It is also coherent with the fact that uniform magnetic 
field components dominate over turbulent components in OMC-1 \citep{hil09} and that an external field 
could be wrapping the OMC-1 filament \citep{val07}. 

This research is supported by the Conseil de recherche en sciences naturelles et en g\'enie du Canada.
FP thanks the Funda\c{c}\~ao de Amparo \`a Pesquisa do Estado de S\~ao Paulo 
(FAPESP grant number 2007/56302-4) for its support during the second phase of the present research.  
The authors thank B. Malenfant and G. Turcotte for their helpful and friendly support during 
observations at Mont-M\'egantic Observatory. 
This work made use of the SIMBAD database at the Canadian Astronomy Data Center, 
which is operated by the Dominion Astrophysical
Observatory for the National Research Council's Herzberg Institute of Astrophysics.

%------- P versus Av -------------------------------

\clearpage

%------- Large Scale visible results ----------------------------

\clearpage

\clearpage

\clearpage

%------- OMC2 region results ---------------------------- 

\clearpage

\clearpage

\clearpage

%------- BN region results ----------------------------

\clearpage

\clearpage

\clearpage

\clearpage

%------ Large scale comparison ---------------------------------------------

\clearpage

%------ omc-1 comparison ---------------------------------------------

\clearpage
\bibliographystyle{apj}
\bibliography{apj-jour,clemson}

%------ BN comparison ---------------------------------------------

\clearpage

\clearpage
\bibliographystyle{apj}
\bibliography{apj-jour,clemson}

%------ P(K) and tau(K) vs offsets in and around BN -------------

\clearpage

\clearpage

%------ omc-23 comparison ---------------------------------------------

\clearpage

\clearpage

%------ omc-4 comparison ---------------------------------------------

\clearpage

%------ OMC-1 magnetic field rotation - schematic view ---------------------------------------------

\clearpage

\clearpage

\begin{deluxetable}{rcllrrrrcrrc} 
\tablewidth{0pt}
\tabletypesize{\scriptsize}
\tablecaption{New visible polarization data from the OMM. \label{OMMDATA}}
\tablehead{
  \colhead{Parenago}                         & \colhead{HD}                          &
  \colhead{$\alpha (2000)$}                  & \colhead{$\delta (2000)$}             & 
  \colhead{ $P_{R} $}                            & \colhead{$\sigma_{P_{R}}$}                 &
  \colhead{$\theta_{R}$}                         & \colhead{$\sigma_{\theta_{R}}$}           &
  \colhead{$\frac{P}{\sigma_{P}}>3$}         & \colhead{ $A_{\rm V}$}                &  
  \colhead{Distance}                         & \colhead{SC$^{(a)}$}                  \\
  \colhead{    }                             & \colhead{ }                        &
  \colhead{($^{h}$ $^{\rm mn}$ $^{\rm s}$)}  & \colhead{($^{\circ}$ $'$ $''$)}               & 
  \colhead{ ($\%$)}                          & \colhead{($\%$)     }                 &
  \colhead{$(^{\circ})$}                         & \colhead{($^{\circ}$)}                    &
  \colhead{}                                 & \colhead{ (Mag.)}                     & 
  \colhead{ (pc)}                            & \colhead{}                            \\
  \colhead{(1)}                                & \colhead{(2)}                           &
  \colhead{(3)}                                & \colhead{(4)}                           & 
  \colhead{(5)}                                & \colhead{(6)}                           &
  \colhead{(7)}                                & \colhead{(8)}                           &
  \colhead{(9)}                                & \colhead{(10)}                          &  
  \colhead{(11)}                               & \colhead{(12)}                          }
\startdata
P1374 &  ...    &05 34 11.13&-05 22 54.6& 0.29& 0.10& 59.3&10.3&0&?   &  ?   &0\\ 
P1539 &  ...    &05 34 40.04&-05 10 07.1& 5.34& 0.13& 95.7& 0.7&1&1.94& ?    &2\\
P1587 &  ...    &05 34 45.20&-05 25 04.1& 2.53& 0.17& 91.7& 1.9&1&?   &  ?   &?\\ 
P1659 &  ...    &05 34 55.98&-05 23 13.1& 0.78& 0.12& 69.4& 4.3&1&?   &  ?   &2\\
P1665b&  ...    &05 34 51.5 &-05 13.5   & 2.14& 0.41& 49.1& 5.4&1&?   &  ?   &2\\
P1665 &  ...    &05 34 56.82&-05 11 33.5& 0.26& 0.21& 40.4&23.1&0&2.15&   ?  &0\\
P1684 &  ...    &05 35 00.11&-05 23 01.9& 0.27& 0.17&115.7&17.4&0&?   &   ?  &0\\
P1746 &  ...    &05 35 05.64&-05 25 19.5& 1.11& 0.13& 64.6& 3.5&1&  ? &  ?   &?\\ 
P1763 &  ...    &05 35 07.53&-05 36 19.3& 0.51& 0.14& 29.1& 7.6&1& ?  &  ?   &?\\
P1773 &  ...    &05 35 09.07&-05 29 59.2& 0.30& 0.14& 59.6&13.4&0&?   &   ?  &0\\
P1775 &  ...    &05 35 07.5 &-06 09 59.0& 0.77& 0.08& 52.7& 3.1&1&1.74&  ?   &2\\
P1786 &  ...    &05 35 10.34&-05 38 24.9& 5.26& 0.11& 91.7& 0.6&1&?   &   ?  &?\\
P1802 &  ...    &05 35 11.14&-05 36 51.2& 1.47& 0.17& 22.6& 3.3&1&?   & ?    &?\\
P1810 &  ...    &05 35 12.23&-05 36 39.4& 0.15& 0.15&138.3&28.2&0&?   & ?    &0\\
P1827 &  ...    &05 35 13.61&-05 32 51.1& 0.53& 0.13& 20.2& 6.9&1&?   &  ?   &?\\
P1828 &  ...    &05 35 12.83&-05 39 34.3& 0.19& 0.17&142.9&26.4&0&?   &  ?   &0\\
P1835 &  ...    &05 35 15.79&-05 00 33.5& 0.49& 0.21&138.4&12.3&0&1.65&  ?   &0\\
P1874 &  ...    &05 35 15.80&-05 33 12.3& 0.93& 0.16&178.2& 4.9&1&?   &  ?   &?\\
P1876 &  ...    &05 35 14.68&-05 35 14.7& 0.61& 0.12&124.5& 5.4&1&?   &  ?   &?\\
P1897 &  ...    &05 35 16.65&-05 32 47.4& 0.92& 0.21& 10.9& 6.5&1&?   &  ?   &?\\
P1898 &  ...    &05 35 15.92&-05 39 14.8& 0.43& 0.14& 37.5& 9.1&1&?   &  ?   &?\\
P1977 &  ...    &05 35 21.7 &-05 34 46.8& 0.67& 0.13&154.9& 5.4&1&?   & ?    &?\\
P2005 &  ...    &05 35 25.24&-05 09 27.6& 0.21& 0.13& 57.9&17.5&0&1.17&   ?  &0\\
P2006 &  ...    &05 35 25.63&-05 09 49.2& 0.31& 0.13& 56.9&11.5&0&1.17&   ?  &0\\
P2007 &  ...    &05 35 25.42&-05 10 48.0& 0.84& 0.24&176.7& 8.2&0&?   &   ?  &?\\
P2029 &  ...    &05 35 26.75&-05 11 07.1& 1.49& 0.18& 25.9& 3.5&1&?   &   ?  &?\\
P2030 &  ...    &05 35 26.88&-05 13 14.0& 0.14& 0.15& 84.3&29.3&0&?   &   ?  &0\\
P2048 &  ...    &05 35 27.43&-05 35 19.4& 0.91& 0.35& 36.6&10.9&0&?   & ?    &0\\
P2057 &  ...    &05 35 28.92&-05 06 03.5& 0.29& 0.22&178.4&22.1&0&0.90&  ?   &0\\
P2068 &  ...    &05 35 31.00&-05 04 14.8& 0.14& 0.13& 69.8&27.8&0&1.12&   ?  &0\\
P2119 &  ...    &05 35 35.98&-05 12 25.1& 0.13& 0.13& 22.0&29.3&0&?   &  ?   &0\\ 
P2143 &  ...    &05 35 38.96&-05 08 55.9& 0.34& 0.14& 36.7&11.8&0&0.98&  ?   &0\\
P2164 &  ...    &05 35 42.91&-05 20 13.3& 0.07& 0.11& 43.8&44.8&0&?   &  ?   &0\\ 
P2173 &  ...    &05 35 42.78&-05 11 54.7& 0.68& 0.19& 57.5& 7.9&1&?   &  ?   &?\\
P2174 &  ...    &05 35 42.94&-05 13 45.2&v $^{(b)}$&&     &    & &?   &  ?   &0\\ 
P2208 &  ...    &05 35 46.89&-05 18 01.6&11.85& 0.16& 64.1& 0.4&1&?   &  ?   &2\\  
P2216 &  ...    &05 35 47.68&-05 10 30.3&00.28& 0.16& 16.2&15.9&0&1.07&  ?   &0\\
P2217 &  ...    &05 35 47.40&-05 16 57.8& 7.87& 0.15& 65.1& 0.5&1&?   &  ?   &2\\
P2244 &  ...    &05 35 51.64&-05 08 09.2& 2.98& 0.19& 75.7& 1.8&1&0.76&  ?   &0\\
P2252 &  ...    &05 35 52.63&-05 06 56.7& 0.27& 0.14& 78.2&14.7&0&0.76&  ?   &0\\
P2317 &  ...    &05 36 04.36&-05 07 14.4& 1.95& 0.13& 78.9& 2.0&1&1.26&   ?  &?\\ 
\enddata
\tablenotetext{(a)}{SC: Selection Code, indicates the reliability of polarization data for probing magnetic fields. SC $=$ 1 or 2 for reliable data. SC $=$ ? for data considered as less reliable, and SC $=$ 0 for data considered as non reliable. 
See section \ref{pvisclass} for details.} 
\tablenotetext{(b)}{v: apparently variable star.}
\end{deluxetable}

\clearpage

\begin{deluxetable}{lrrrrrrrcrrc} 
\tablewidth{0pt}
\tabletypesize{\scriptsize}
\tablecaption{Visible polarization data from stars observed at OMM, and from \citet{bre76}
and/or from the \citet{hei00} catalog. \label{PVISDATA}}
\tablehead{
  \colhead{Parenago}                         & \colhead{HD}                          &
  \colhead{$\alpha (2000)$}                  & \colhead{$\delta (2000)$}             & 
  \colhead{ $P $}                            & \colhead{$\sigma_ P$}                 &
  \colhead{$\theta$}                         & \colhead{$\sigma_{\theta}$}           &
  \colhead{$\frac{P}{\sigma_{P}}>3$}         & \colhead{ $A_{\rm V}$}                &  
  \colhead{Distance}                         & \colhead{SC$^{(a)}$}                  \\
  \colhead{ }                                & \colhead{ }                        &
  \colhead{($^{h}$ $^{\rm mn}$ $^{\rm s}$)}  & \colhead{($^{\circ}$ $'$ $''$)}               & 
  \colhead{ ($\%$)}                          & \colhead{($\%$)     }                 &
  \colhead{$(^{\circ})$}                         & \colhead{($^{\circ}$)}                    &
  \colhead{}                                 & \colhead{ (Mag.)}                     & 
  \colhead{ (pc)}                            & \colhead{}                            \\
  \colhead{(1)}                                & \colhead{(2)}                           &
  \colhead{(3)}                                & \colhead{(4)}                           & 
  \colhead{(5)}                                & \colhead{(6)}                           &
  \colhead{(7)}                                & \colhead{(8)}                           &
  \colhead{(9)}                                & \colhead{(10)}                          &  
  \colhead{(11)}                               & \colhead{(12)}                          }
\startdata
P1391$^{(\rm ISM)}$ &      &05 34 15.20&-05 11 49.6& 0.38$^{(M)}$& 0.12& 53.3&09.4&1&?   &  ?   &?\\
...   &      &...        &...        & 0.07$^{(B)}$& 0.12& 89.0&48.9&0& ...&...   &0\\
P1455$^{(\rm ISM)}$ &      &05 34 24.83&-05 22 05.1& 0.15$^{(M)}$& 0.12& 41.3&23.2&0&  ? & ?    &0\\ 
...   &      & ...       &...        & 0.04$^{(B)}$& 0.06& 66.0&42.8&0&... &  ... &...\\ 
P1507$^{(\rm ISM)}$ &      &05 34 34.47&-05 03 06.9& 0.69$^{(M)}$& 0.11& 96.8& 4.5&1&1.14&  ?   &1\\ 
...   &      &...        &...        & 0.55$^{(B)}$& 0.05&102.0& 2.5&1&... & ...  &...\\
P1540$^{(\rm ICD)}$ &      &05 34 39.79&-05 24 25.7& 0.79$^{(M)}$& 0.11& 79.3& 4.1&1&   ?&  ?   &1\\
...   &      &...        &...        & 1.11$^{(B)}$& 0.10& 85.0& 2.6&1& ...&...   &...\\
P1562$^{(\rm ISM)}$ &36899 &05 34 42.28&-05 07 14.6& 0.35$^{(M)}$& 0.12& 50.2& 9.9&0&1.59& 602.9&0\\ 
...   &  ... &...        & ...       & 0.89$^{(H)}$& 0.03&171.0& 1.0&1& ...& ...  &...\\ 
...   & ...  & ...       & ...       & 0.27$^{(B)}$& 0.08& 88.0& 8.4&1& ...&...   &...\\ 
P1575$^{(\rm ISM)}$ &      &05 34 43.54&-05 18 27.2& 0.21$^{(M)}$& 0.16& 86.0&21.6&0&?   &  ?   &0\\ 
...   &      &...        & ....      & 0.18$^{(B)}$& 0.09& 89.0&14.3&0& ...&...   &...\\ 
P1605$^{(\rm ICD)}$ &36917 &05 34 46.98&-05 34 14.6& 1.06$^{(M)}$& 0.11& 44.6& 3.0&1&0.62& 239.9&1\\
...   & ...  & ...       & ...       & 0.97$^{(H)}$& 0.03& 43.0& 0.9&1& ...& ...  &...\\
...   & ...  & ...       & ...       & 0.84$^{(B)}$& 0.05& 42.0& 1.7&1& ...&...   &...\\
P1623$^{(\rm ICD)}$ &      &05 34 49.99&-05 18 44.7& 2.33$^{(M)}$& 0.16& 53.6& 1.9&1&?   &  ?   &1\\
...   &      & ...       & ...       & 1.83$^{(B)}$& 0.04& 52.0& 0.6&1& ...& ...  &...\\
P1660$^{(\rm ISM)}$ &36939 &05 34 55.29&-05 30 22.1& 0.49$^{(M)}$& 0.10& 61.2& 5.9&1&0.31& 524.8&1\\
 ...  & ...  & ...       & ...       & 0.71$^{(H)}$& 0.03& 65.0& 1.3&1&... & ...  &...\\
 ...  & ...  & ...       & ...       & 0.53$^{(B)}$& 0.07& 53.0& 3.8&1&... & ...  &...\\
P1683$^{(\rm ICD)}$ &      &05 35 00.73&-05 05 11.0& 1.42$^{(M)}$& 0.13&108.0& 2.6&1&1.48&  ?   &1\\
...   &      & ...       & ...       & 1.36$^{(B)}$& 0.07&117.0& 1.5&1& ...&...   &...\\ 
P1685$^{(\rm ICD)}$ &      &05 35 00.14&-05 25 16.3& 1.41$^{(M)}$& 0.10& 78.7& 2.1&1&?   &   ?  &1\\ 
 ...  &      &    ...    & ...       & 1.48$^{(B)}$& 0.07& 67.0& 1.4&1& ...& ...  &...\\
P1712$^{(\rm ICD)}$ &      &05 35 03.62&-05 05 44.5& 3.44$^{(M)}$& 0.12&100.2& 1.0&1&1.48&  ?   &1\\ 
 ...  & ...  & ...       & ...       & 3.73$^{(B)}$& 0.09&100.0& 0.7&1& ...& ...  &...\\
P1724$^{(\rm ICD)}$ &      &05 35 04.3 &-05 08 12.8& 1.11$^{(M)}$& 0.10& 83.0& 2.6&1&1.51&  ?   &1\\
 ...  & ...  & ...       & ...       & 1.50$^{(B)}$& 0.04& 79.0& 0.8&1&... & ...  &...\\
P1744$^{(\rm ISM)}$ &36981 &05 35 06.20&-05 12 15.9& 0.14$^{(M)}$& 0.10& 49.6&20.8&0&  ? & ?    &0\\
 ...  & ...  & ...       & ...       & 0.07$^{(H)}$& 0.03& 84.0&13.0&0& ...& ...  &0\\
 ...  & ...  & ...       & ...       & 0.36$^{(B)}$& 0.07& 81.0& 5.5&1& ...& ...  &?\\
P1885$^{(\rm ICD)}$ &      &05 35 16.99&-05 21 45.6& 1.80$^{(M)}$& 0.13&107.5& 2.0&1&?   &  ?   &1\\ 
 ...  & ...  & ...       & ...       & 0.98$^{(B)}$& 0.08&105.0& 2.3&1& ?  &  ?   &...\\
...   & ...  & ...       & ...       & 0.06$^{(B)}$& 0.15&162.0&71.2&0&... & ...  &...\\ 
P1905$^{(\rm ISM)}$ &37019 &05 35 18.20&-05 03 54.9& v $^{(b)}$$^{(M)}$&     &     &    & &1.43& 436.5&0\\
...   & ...  & ...       & ...       & 0.35$^{(H)}$& 0.03& 96.0& 2.6&1& ...& ...  &...\\
...   & ...  & ...       & ...       & 0.28$^{(B)}$& 0.08& 48.0& 8.2&1&... & ...  &...\\
P1950$^{(\rm ISM)}$ &      &05 35 23.4 & -04 53 01 & 0.63$^{(M)}$& 0.15& 40.1& 6.7&1&?   &  ?   &0\\
  ... &      &    ...    & ...       & 0.29$^{(B)}$& 0.08& 70.0& 7.9&1& ...&...   &...\\
P1953$^{(\rm ISM)}$ &      &05 35 21.23&-05 09 16.1& 0.24$^{(M)}$& 0.13& 22.3&15.0&0&?   & ?    &0\\
  ... &      &    ...    &   ...     & 0.26$^{(B)}$& 0.10& 66.0&11.0&0& ...&...   &...\\ 
P1956$^{(\rm ICD)}$ &      &05 35 20.65&-05 21 44.9& 1.93$^{(M)}$& 0.09& 37.9& 1.4&1&?   & ?    &1\\
...   &      &    ...    &   ...     & 1.30$^{(B)}$& 0.07& 40.0& 1.5&1& ...&...   &...\\ 
P2031$^{(\rm ISM)}$ &37042 &05 35 26.40&-05 25 00.7& 0.37$^{(M)}$& 0.10&140.7& 7.4&1&0.62& 757.0&1\\
  ... & ...  & ...       & ...       & 0.56$^{(H)}$& 0.03&139.0& 1.6&1& ...& ...  &...\\
 ...  & ...  & ...       & ...       & 0.27$^{(B)}$& 0.04&135.0& 4.2&1& ...& ...  &...\\
P2043$^{(\rm ISM)}$ &      &05 35 28.5 &-04 55 03.2& 0.12$^{(M)}$& 0.15& 68.0&34.2&0&?   &  ?   &0\\
 ...  &      & ...       & ...       & 0.34$^{(B)}$& 0.17& 60.0&14.2&0& ...& ...  &...\\ 
P2065$^{(\rm ISM)}$ &37059 &05 35 31.16&-04 54 15.4& 0.21$^{(M)}$& 0.10& 48.9&13.3&0&0.31& 602.6&0\\
 ...  & ...  & ...       & ...       & 0.75$^{(H)}$& 0.03& 86.0& 1.2&1& ...& ...  &...\\
...   & ...  & ...       & ...       & 0.08$^{(B)}$& 0.05&118.0&17.8&0& ...& ...  &...\\ 
P2074$^{(\rm ICD)}$ &37061 &05 35 31.36&-05 16 02.6& 1.65$^{(M)}$& 0.08& 59.4& 1.4&1&?   &   ?  &1\\
 ...  & ...  & ...       & ...       & 1.37$^{(B)}$& 0.05& 62.0& 1.0&1& ...& ...  &...\\ 
P2102$^{(\rm ISM)}$ &37060 &05 35 34.28&-05 06 21.2& 0.16$^{(M)}$& 0.11& 71.1&19.4&0&0.82&  ?   &0\\
 ...  & ...  & ...       & ...       & 0.30$^{(B)}$& 0.14& 74.0&13.3&0&... &...   &...\\ 
P2284$^{(\rm ISM)}$ &37114 &05 35 58.53&-05 22 31.6& 0.18$^{(M)}$& 0.10& 42.5&15.0&0&0.31& 559.0&0\\
 ...  & ...  & ...       & ...       & 0.39$^{(H)}$& 0.03& 88.0& 2.3&1& ...& ...  &1\\
 ...  & ...  & ...       & ...       & 0.31$^{(B)}$& 0.10& 69.0& 9.2&1& ...& ...  &1\\
P2290$^{(\rm ISM)}$ &294266&05 36 00.70&-04 57 57.1& 0.18$^{(M)}$& 0.12& 25.8&18.5&0&1.82& ?    &0\\
 ...  & ...  & ...       & ...       & 0.13$^{(B)}$& 0.08&175.0&17.5&0& ...& ...  &...\\
P2342$^{(\rm ISM)}$ &37142 &05 36 11.02&-05 03 41.5& 0.35$^{(M)}$& 0.11& 54.8& 8.9&1&1.64&  ?   &2\\
 ...  & ...  & ...       & ...       & 0.15$^{(B)}$& 0.08& 45.0&15.2&0& ...& ...  &0\\ 
P2387$^{(\rm ISM)}$ &37174 &05 36 27.19&-05 24 31.3& 0.29$^{(M)}$& 0.09& 60.3& 9.0&1&2.05& 524.8&0\\
 ...  & ...  & ...       & ...       & 0.33$^{(H)}$& 0.03&156.0& 2.8&1&... & ...  &...\\
...   & ...  & ...       & ...       & 0.06$^{(B)}$& 0.15&162.0&71.2&0&... & ...  &...\\ 

\enddata
\tablenotetext{(\rm ISM)}{Polarization mainly produced by ISM dust \citep[see][]{bre76}.}
\tablenotetext{(\rm ICD)}{Polarization mainly produced by ICD \citep[see][]{bre76}.}
\tablenotetext{(a)}{SC: Selection Code, indicates the reliability of polarization data for probing magnetic fields. SC $=$ 1 or 2 for reliable data. SC $=$ ? for data considered as less reliable, and SC $=$ 0 for data considered as non reliable. 
See section \ref{pvisclass} for details.}
\tablenotetext{(b)}{v: apparently variable star.}
\tablenotetext{(B)}{Data from \citet{bre76}, see also \citet{bre77} where the origin of the 
polarization is discussed for some stars. The effective wavelength of observations lies 
between B and V effective wavelength.}
\tablenotetext{(H)}{Data from \citet{hei00}; $\lambda$ is not defined precisely.}
\tablenotetext{(M)}{Data from Mont-M\'egantic Observatory.}
\end{deluxetable}

\clearpage

\clearpage

\begin{deluxetable}{lrrrrrrrcrrc} 
\tablewidth{0pt}
\tabletypesize{\scriptsize}
\tablecaption{Supplementary visible polarization data from \citet{bre76} and 
from the \citet{hei00} catalog. \label{OTHERDATA}}
\tablehead{
  \colhead{Parenago}                         & \colhead{HD}                          &
  \colhead{$\alpha (2000)$}                  & \colhead{$\delta (2000)$}             & 
  \colhead{ $P $}                            & \colhead{$\sigma_ P$}                 &
  \colhead{$\theta$}                         & \colhead{$\sigma_{\theta}$}           &
  \colhead{$\frac{P}{\sigma_{P}}>3$}         & \colhead{ $A_{\rm V}$}                &  
  \colhead{distance}                         & \colhead{SC$^{(a)}$}                  \\
  \colhead{ }                             & \colhead{ }                        &
  \colhead{($^{h}$ $^{\rm mn}$ $^{\rm s}$)}  & \colhead{($^{\circ}$ $'$ $''$)}               & 
  \colhead{ ($\%$)}                          & \colhead{($\%$)     }                 &
  \colhead{$(^{\circ})$}                         & \colhead{($^{\circ}$)}                    &
  \colhead{}                                 & \colhead{ (Mag.)}                     & 
  \colhead{ (pc)}                            & \colhead{}                            \\
  \colhead{(1)}                                & \colhead{(2)}                           &
  \colhead{(3)}                                & \colhead{(4)}                           & 
  \colhead{(5)}                                & \colhead{(6)}                           &
  \colhead{(7)}                                & \colhead{(8)}                           &
  \colhead{(9)}                                & \colhead{(10)}                          &  
  \colhead{(11)}                               & \colhead{(12)}                          }
\startdata
P1036$^{(\rm ICD)}$ &      &05 32 55.89&-04 32 42.7& 2.33$^{(B)}$& 0.10& 95.0& 1.2&1&0.49&  ?   &2\\
P1044$^{(\rm ICD)}$ &36629 &05 32 57.08&-04 33 59.3& 1.96$^{(B)}$& 0.02& 97.0& 0.3&1&0.64& 719.0&1\\
...   &  ... &        ...&        ...& 1.84$^{(H)}$& 0.05& 95.8& 0.7&1&... & ...  &...\\
P1049$^{(\rm ICD)}$ &      &05 32 58.00&-04 30.8   & 2.34$^{(B)}$& 0.14&108.0& 1.7&1&0.49&   ?  &1\\
P1073$^{(\rm ICD)}$ &      &05 33 00.0 &-06 10.1   & 3.97$^{(B)}$& 0.29& 36.0& 2.1&1&0.20&  ?   &?\\
P1097$^{(\rm ISM)}$ &36655 &05 33 07.48&-05 20 26.1& 0.50$^{(H)}$& 0.03& 47.0& 1.8&1&1.18& 478.6&0\\
...   &...   &...        &...        & 0.37$^{(B)}$& 0.05& 76.0& 3.9&1&... & ...  &...\\
P1117$^{(\rm ISM)}$ &36671 &05 33 14.03&-04 38 07.2& 0.39$^{(H)}$& 0.03& 82.0& 2.3&1&1.22& 316.2&1\\
...   &...   &...        &...        & 0.49$^{(B)}$& 0.08&100.0& 4.7&1&... & ...  &...\\
P1122$^{(\rm ICD)}$ &      &05 33 16.0 &-04 28.2   & 1.41$^{(B)}$& 0.12& 81.0& 2.4&1&0.44&  ?   &2\\
P1150$^{(\rm ICD)}$ &      &05 33 20.07&-05 06 24.3& 5.91$^{(B)}$& 0.06& 97.0& 0.3&1&1.45&  ?   &1\\
P1175$^{(\rm ICD)}$ &      &05 33 24.3 &-05 21 16  & 0.57$^{(B)}$& 0.06& 98.0& 3.0&1&1.47&   ?  &2\\
P1212$^{(\rm ICD)}$ &294224&05 33 34.06&-05 05 01.7& 4.21$^{(B)}$& 0.06& 87.0& 0.4&1&1.39&  ?   &1\\
P1427$^{(\rm ICD)}$ &      &05 34 22.26&-04 57 40.2& 1.03$^{(B)}$& 0.18& 32.0& 5.0&1&1.52& ?    &?\\
P1466$^{(\rm ICD)}$ &      &05 34 24.9 &-06 01 03  & 1.12$^{(B)}$& 0.11& 61.0& 2.8&1&1.29&   ?  &?\\
P1491$^{(\rm ISM)}$ &36865 &05 34 32.43&-04 29 14.8& 0.14$^{(H)}$& 0.03& 68.0& 6.5&1&0.31& 281.0&1\\
...   &...   &...        &...        & 0.17$^{(B)}$& 0.04& 77.0& 6.7&1&... & ...  &...\\
P1526$^{(\rm ICD)}$ &      &05 34 41.0 &-04 17.5   & 1.57$^{(B)}$& 0.06&104.0& 1.1&1&0.54& ?    &2\\
P1546$^{(\rm ISM)}$ &36883 &05 34 43.20&-04 23 31.5& 0.23$^{(H)}$& 0.03& 41.0& 4.0&1&0.26& 521.0&1\\
...   & ...  & ...       & ...       & 0.22$^{(B)}$& 0.10& 48.0&13.0&0& ...& ...  &...\\
P1567$^{(\rm ICD)}$ &      &05 34 44.5 &-04 42 14  & 2.33$^{(B)}$& 0.23& 97.0& 2.8&1&2.10&  ?   &2\\
P1581$^{(\rm ICD)}$ &      &05 34 48   &-04 17.5   & 2.27$^{(B)}$& 0.12&123.0& 1.5&1&0.45&  ?   &2\\
P1628               &36916 &05 34 53.96&-04 06 37.5& 0.27$^{(H)}$& 0.03& 58.0& 3.4&1&0.89& 356.0&2\\
P1654$^{(\rm ICD)}$ &36938 &05 34 56.24&-04 45 57.4& 0.90$^{(H)}$& 0.03& 44.0& 1.0&1&0.62& 478.6&0\\
...   & ...  & ...       & ...       & 0.62$^{(B)}$& 0.10& 22.0& 4.6&1&... &...   &...\\
P1664$^{(\rm ISM)}$ &36936 &05 34 59.00&-04 21 15.4& 0.18$^{(H)}$& 0.03& 48.0& 5.1&1&0.26& 512.0&2\\
P1698$^{(\rm ISM)}$ & 36957&05 35 03.73&-04 23 06.1& 0.19$^{(H)}$& 0.03& 84.0& 4.8&1&0.32& 398.1&0\\
  ... & ...  & ...       & ...       & 0.50$^{(B)}$& 0.07& 55.0& 4.0&1&... & ...  &...\\
P1708$^{(\rm ISM)}$ &36958 &05 35 04.79&-04 43 54.6& 0.94$^{(H)}$& 0.08& 47.7& 2.3&1&0.62& 511.0&1\\
...   & ...  & ...       & ...       & 0.59$^{(B)}$& 0.06& 38.0& 2.9&1&... &...   &...\\
P1716$^{(\rm ISM)}$ &36959 &05 35 01.01&-06 00 33.4& 0.16$^{(H)}$& 0.0 & 82.0& 0.0&1&0.31& 646.0&0\\
 ...  & ...  & ...       & ...       & 0.19$^{(B)}$& 0.04&131.0& 6.0&1&... & ...  &...\\
P1719$^{(\rm ICD)}$ &      &05 35 05.42&-04 43 19.8& 0.73$^{(B)}$& 0.06&116.0& 2.4&1&?   &   ?  &?\\
P1728$^{(\rm ISM)}$ &36960 &05 35 02.68&-06 00 07.3& 0.11$^{(H)}$& 0.0 & 69.0& 0.0&1&0.31& 539.0&1\\
...   & ...  & ...       & ....      & 0.13$^{(B)}$& 0.03& 70.0& 6.6&1& ...& ...  &...\\
P1736$^{(\rm ICD)}$ &      &05 35 37.58&-05 16 44.5& 1.73$^{(B)}$& 0.16&123.0& 2.7&1&  ? & ?    &?\\
P1772$^{(\rm ICD)}$ & 36982&05 35 09.84&-05 27 53.3& 0.55$^{(B)}$& 0.03& 64.0& 1.6&1&1.24&1139.0&1\\
 ...  & .... & ...       & ...       & 1.01$^{(H)}$& 0.02& 56.0& 0.6&1&... & ...  &...\\
P1795$^{(\rm ICD)}$ &36998 &05 35 13.23&-04 37 34.6& 0.27$^{(H)}$& 0.03&169.0& 3.4&1&1.70& 575.4&0\\
 ...  & ...  & ...       & ...       & 0.30$^{(B)}$& 0.05& 63.0& 4.8&1&... & ...  &...\\
P1798$^{(\rm ICD)}$ &294264&05 35 13.35&-04 51 44.9& 2.62$^{(B)}$& 0.08& 77.0& 0.9&1&?   & ?    &1\\
P1799$^{(\rm ICD)}$ &      &05 35 11.64&-05 16 57.8& 5.84$^{(B)}$& 0.12&134.0& 0.6&1&?   & ?    &1\\
P1813$^{(\rm ISM)}$ &37000 &05 35 11.01&-05 55 36.9& 0.35$^{(H)}$& 0.03& 84.0& 2.6&1&0.31& 465.0&?\\
 ...  & ...  & ...       & .. .      & 0.07$^{(B)}$& 0.03& 96.0&12.2&0&... & ...  &0\\
P1854$^{(\rm ICD)}$ &294263&05 35 17.83&-04 41 07.0& 0.63$^{(B)}$& 0.08& 71.0& 3.6&1&?   &  ?   &?\\
P1881$^{(\rm ICD)}$ & 36998&05 35 18.42&-04 40 55.9& 1.03$^{(B)}$& 0.05&107.0& 1.4&1&?   &  ?   &1\\
P1901$^{(\rm ISM)}$ & 37025&05 35 15.75&-06 01 57.9& 0.06$^{(H)}$& 0.03& 99.0&15.2&0&1.60&  ?   &0\\
 ...  & ...  & ...       & ...       & 0.16$^{(B)}$& 0.02& 60.0& 3.6&1& ...&...   &?\\
P1921$^{(\rm ICD)}$ &      &05 35 18.91&-05 16 14.1& 1.41$^{(B)}$& 0.33&169.0& 6.7&1& ?  & ?    &?\\
P1932$^{(\rm ISM)}$ & 37016&05 35 22.32&-04 25 27.6& 0.35$^{(H)}$&  0.0& 55.0& 0.0&1&0.74& 345.0&0\\
...   & ...  & ...       & ...       & 0.36$^{(B)}$& 0.08& 29.0& 6.3&1& ...& ...  &...\\
P1933$^{(\rm ISM)}$ &37017 &-05 35 21.8&-04 29 39.0& 0.25$^{(H)}$& 0.0 & 51.0& 0.0&1&0.96& 630.0&1\\
...   & ...  & ...       & ...       & 0.40$^{(B)}$& 0.08& 49.0& 5.7&1&... & ...  &...\\
P1970$^{(\rm ISM)}$ &37018 &05 35 23.16&-04 50 18.1& 0.47$^{(H)}$& 0.03& 62.0& 2.2&1&0.31& 470.0&0\\
 ...  &      &    ...    &   ...     & 0.31$^{(B)}$& 0.05&151.0& 4.6&1& ...&...   &...\\
P1993$^{(\rm ICD)}$ &37041 &05 35 22.90&-05 24 57.8& 0.79$^{(H)}$& 0.05& 95.8& 1.8&1&0.62& 607.0&1\\
...   & ...  & ...       & ...       & 0.96$^{(B)}$& 0.02& 99.0& 0.6&1& ...&  ... &...\\
P2001$^{(\rm ICD)}$ &      &05 35 23.84&-05 30 47.5& 1.02$^{(B)}$& 0.16&175.0& 4.5&1&?   &   ?  &?\\
P2037$^{(\rm ISM)}$ &37043 &05 35 25.98&-05 54 35.6& 0.11$^{(H)}$& 0.0 & 61.0& 0.0&1&0.31& 451.0&0\\
 ...  & ...  & ...       & ...       & 0.21$^{(B)}$& 0.04&133.0& 5.4&1& ...& ...  &...\\
P2054 &37040 &05 35 31.08&-04 21 50.6& 0.23$^{(H)}$&  0.0& 61.0& 0.0&1&0.95& 500.0&?\\
P2085$^{(\rm ISM)}$ &37062 &05 35 31.44&-05 25 16.4& 0.41$^{(H)}$& 0.03&162.0& 2.2&1&1.86& 689.0&0\\
 ...  & ...  & ...       & ...       & 0.14$^{(B)}$& 0.06& 47.0&12.2&0& ...& ...  &...\\
P2086$^{(\rm ICD)}$ &      &05 35 31.25&-05 33 11.8& 1.63$^{(B)}$& 0.08& 77.0& 1.4&1&?   &  ?   &1\\
P2185$^{(\rm ICD)}$ &      &05 35 43.22&-05 36 27.6& 0.90$^{(B)}$& 0.15&114.0& 4.8&1&?   &  ?   &?\\
P2233$^{(\rm ICD)}$ &      &05 35 50.7 &-04 51 11  & 3.09$^{(B)}$& 0.11&123.0& 1.0&1&1.71&  ?   &?\\
P2248$^{(\rm ICD)}$ &      &05 35 49.83&-05 40 27.6& 3.13$^{(B)}$& 0.08& 90.0& 0.7&1&?   &  ?   &?\\
P2261$^{(\rm ICD)}$ &      &05 35 54.9 &-04 58 07.0& 1.05$^{(B)}$& 0.12& 40.0& 3.3&1&2.17&  ?   &?\\
P2271$^{(\rm ISM)}$ &37115 &05 35 54.08&-05 37 42.3& 0.37$^{(H)}$& 0.03& 89.0& 2.5&1&0.31& 304.0&1\\
...   & ...  & ...       & ...       & 0.31$^{(B)}$& 0.07&105.0& 6.4&1& ...& ...  &...\\
P2302$^{(\rm ICD)}$ &37130 &05 36 03.57&-04 45 07.6& 1.02$^{(B)}$& 0.12&136.0& 3.4&1&1.94&   ?  &2\\
P2310$^{(\rm ICD)}$ &      &05 36 04.3 &-04 44 39.0& 3.83$^{(B)}$& 0.09&133.0& 0.7&1&1.94&   ?  &2\\
P2314$^{(\rm ISM)}$ & 37129&05 36 06.26&-04 25 32.8& 0.32$^{(H)}$& 0.03& 64.6& 2.6&1&1.05& 631.0&1\\
 ...  & ...  & ...       & ...       & 0.23$^{(B)}$& 0.04& 53.0& 5.0&1&... & ...  &...\\
P2368$^{(\rm ICD)}$ &      &05 36 16.98&-05 11 42.9& 4.69$^{(B)}$& 0.50& 20.0& 3.0&1&1.62& ?    &?\\
P2425$^{(\rm ICD)}$ &      &05 36 38.7 &-05 28 22  & 3.47$^{(B)}$& 0.06& 92.0& 0.5&1&2.02&  ?   &1\\
P2448$^{(\rm ICD)}$ &      &05 36 53.63&-04 39 53.5& 2.13$^{(B)}$& 0.09&155.0& 1.2&1&1.73&  ?   &?\\
P2467$^{(\rm ICD)}$ &      &05 37 02.0 &-04 12.3   & 1.62$^{(B)}$& 0.15& 73.0& 2.6&1&0.61&  ?   &?\\
P2471$^{(\rm ICD)}$ &      &05 37 03.14&-04 37 02.5& 0.93$^{(B)}$& 0.17& 44.0& 5.2&1&1.53&  ?   &?\\
P2500$^{(\rm ICD)}$ &      &05 37 13.43&-05 40 26.9& 1.79$^{(B)}$& 0.02& 87.0& 0.3&1&2.06&  ?   &2\\
P2519$^{(\rm ICD)}$ &      &05 37 21.0 &-04 40.4   & 2.39$^{(B)}$& 0.08&155.0& 1.0&1&1.91&  ?   &1\\
P2579$^{(\rm ICD)}$ &      &05 37 41.41&-04 51 04.1& 1.92$^{(B)}$& 0.11&114.0& 1.6&1&1.38&  ?   &?\\
P2602$^{(\rm ICD)}$ &37356 &05 37 53.39&-04 48 50.5& 1.45$^{(H)}$& 0.06& 70.2& 1.2&1&1.08& 489.0&1\\
 ...  & ...  & ...       & ...       & 1.51$^{(B)}$& 0.03& 75.0& 0.6&1& ...& ...  &...\\
P2609$^{(\rm ICD)}$ &294295&05 37 55.86&-04 47 18.0& 0.75$^{(B)}$& 0.01& 76.0& 0.4&1&1.08&   ?  &2\\
P2653 &      &05 38 14.51&-05 25 13.3& 0.23$^{(H)}$& 0.03& 57.0& 4.0&1&2.10& 478.6&?\\
P2701 &37469 &05 38 39.46&-04 40 48.8& 2.22$^{(H)}$& 0.03&108.0& 0.4&1&1.02& 501.2&?\\
P2758 &      &05 39 02.4 &-05 11 40.1& 0.31$^{(H)}$& 0.03& 39.0& 2.9&1&1.33& 506.0&?\\
\enddata
\tablenotetext{(\rm ISM)}{Polarization mainly produced by ISM dust \citep[see][]{bre76}.}
\tablenotetext{(\rm ICD)}{Polarization mainly produced by ICD \citep[see][]{bre76}.}
\tablenotetext{(a)}{SC: Selection Code, indicates the reliability of polarization data for probing magnetic fields. SC $=$ 1 or 2 for reliable data. SC $=$ ? for data considered as less reliable, and SC $=$ 0 for data considered as non reliable.
See section \ref{pvisclass} for details.}
\tablenotetext{(B)}{Data from \citet{bre76}, see also \citet{bre77} where the origin of the 
polarization is discussed for some stars.}
\tablenotetext{(H)}{Data from \citet{hei00}.}
\end{deluxetable}

\clearpage

\begin{deluxetable}{lcccccccccccccccccc} 
\tablewidth{0pt}
\rotate
\tabletypesize{\scriptsize}
\tablecaption{Visible polarization data in OMC-2 from Mount Lemmon. \label{OMC2DATA}}
\tablehead{
  \colhead{Star$^{(a)}$}                             & 
  \colhead{$\alpha(2000)$}                   & \colhead{$\delta(2000)$}              &
  \colhead{$P_{V}$}                          & \colhead{$\sigma_{P_{V}}$}            &
  \colhead{$\theta_{V}$}                     & \colhead{ $\sigma_{\theta_{V}} $}     &         
  \colhead{$P_{R}$}                          & \colhead{$\sigma_{P_{R}}$}            &
  \colhead{$\theta_{R}$}                     & \colhead{ $\sigma_{\theta_{R}} $}     &         
  \colhead{$P_{I}$}                          & \colhead{$\sigma_{P_{I}}$}            &
  \colhead{$\theta_{I}$}                     & \colhead{ $\sigma_{\theta_{I}} $}     &         
  \colhead{$A_{\rm V}$}                      & 
  \colhead{SC$^{(b)}$}                      \\
  \colhead{}                                 & 
  \colhead{($^{h}$ $^{\rm mn}$ $^{\rm s}$)}  & \colhead{($^{o}$ $'$ $''$)}               & 
  \colhead{$(\%)$}                           & \colhead{$(\%)$}                      & 
  \colhead{$^{(\circ)}$}                         & \colhead{$^{(\circ)}$}                    &
  \colhead{$(\%)$}                           & \colhead{$(\%)$}                      & 
  \colhead{$^{(\circ)}$}                         & \colhead{$^{(\circ)}$}                    &
  \colhead{$(\%)$}                           & \colhead{$(\%)$}                      & 
  \colhead{$^{(\circ)}$}                         & \colhead{$^{(\circ)}$}                    &
  \colhead{(Mag.)}                           & 
  \colhead{}                                 \\
  \colhead{(1)}                                 & 
 \colhead{(2)}                                 & \colhead{(3)}               & 
  \colhead{(4)}                                 & \colhead{(5)}                      & 
  \colhead{(6)}                                 & \colhead{(7)}                    &
  \colhead{(8)}                                 & \colhead{(9)}                      & 
  \colhead{(10)}                                & \colhead{(11)}                    &
  \colhead{(12)}                                & \colhead{(13)}                      & 
  \colhead{(14)}                                & \colhead{(15)}                    &
  \colhead{(16)}                                & 
  \colhead{(17)}                                 }
\startdata
26 & 5 35 03.8 & -5 08 16.8 & 1.38 & 0.12 & 80.0 & 2.5 & 1.30 & 0.06 & 81.0 & 1.3& 1.03 & 0.06 &   1.7 & 1.7& 1.51 & ?\\
76 & 5 35 21.3 & -5 12 09.1 & 0.86 & 0.13 & 18.3 & 4.3 & 0.74 & 0.08 & 13.1 & 3.1& 0.65 & 0.08 &  14.5 & 3.5& ?    & 1\\
85 & 5 35 22.3 & -5 09 11.2 & ...  &  ... &  ... & ... & 0.45 & 0.14 & 63.8 & 8.9& ...  & ...  &  ...  &... & 1.14 & ?\\
111& 5 35 25.6 & -5 09 48.4 & ...  & ...  &  ... & ... & 0.88 & 0.11 & 84.5 & 3.6& 0.69 & 0.21 &  68.0 & 8.7& 1.17 & 1\\
112& 5 35 26.1 & -5 08 40.4 & 0.42 & 0.13 & 58.7 & 8.9 & 0.18 & 0.06 & 79.7 & 9.5& ...  & ...  &  ...  &... & 1.17 & 1\\ 
116& 5 35 26.9 & -5 11 04.5 & ...  & ...  &  ... & ... & 3.57 & 0.18 & 49.3 & 1.4& 3.33 & 0.16 &  52.0 & 1.4&  ?   & 1\\ 
117& 5 35 27.0 & -5 13 08.5 & ...  & ...  &  ... & ... & 0.31 & 0.09 & 48.9 & 8.3& ...  & ...  &  ...  &... & ?    & ?\\
139& 5 35 31.9 & -5 09 26.8 & 0.51 & 0.14 & 74.3 & 7.9 & 0.54 & 0.12 & 60.7 & 6.3& 0.68 & 0.11 &  52.5 & 4.6& 0.98 & 1\\
175& 5 35 40.7 & -5 08 59.5 & ...  & ...  &  ... & ... & 8.83 & 0.34 & 73.2 & 1.1& 8.20 & 0.26 &  73.8 & 0.9& 0.98 & 0\\
183& 5 35 43.0 & -5 11 47.7 & ...  & ...  &  ... & ... & 0.54 & 0.11 & 83.1 & 5.8& ...  & ...  &  ...  &... &?     & ?\\
191& 5 35 44.9 & -5 07 16.8 & ...  & ...  &  ... & ... & 0.20 & 0.06 & 29.8 & 8.6& ...  & ...  &  ...  &... & 0.65 & ?\\
198& 5 35 47.6 & -5 10 24.0 & ...  & ...  &  ... & ... & 0.35 & 0.12 & 76.7 & 9.8& 0.34 & 0.12 &  73.4 &10.1& 1.07 & 1\\
207& 5 35 51.0 & -5 09 26.2 & ...  & ...  &  ... & ... & 8.04 & 0.33 & 74.1 & 1.2& 3.80 & 0.09 &  74.1 & 0.7& 1.06 & 0\\ 
217& 5 35 56.2 & -5 08 58.6 & ...  & ...  &  ... & ... & 4.69 & 0.37 & 97.4 & 2.3& ...  & ...  &  ...  &... & 1.06 & 0\\
\enddata
\tablenotetext{(a)}{Numbers are from \citet{jon94}}
\tablenotetext{(b)}{SC: Selection Code, indicates the reliability of polarization data for probing magnetic fields. SC $=$ 1 or 2 for reliable data. SC $=$ ? for data considered as less reliable, and SC $=$ 0 for data considered as non reliable.  
See section \ref{pvisclass} for details.}
\end{deluxetable}

\clearpage

\begin{deluxetable}{lccccccccccrc}
\tablewidth{0pt}
\tabletypesize{\scriptsize}
\tablecaption{IR polarization data in the BN region. \label{BNDATA}}
\tablehead{
  \colhead{Star}                             & 
  \colhead{$\alpha(2000)$}                   & \colhead{$\delta(2000)$}              &
  \colhead{$P_{K}$$^{(a)}$}                  & \colhead{$\theta_{K}$}                & 
  \colhead{ $\sigma_{\theta_{K}} $}          & \colhead{$H-K$}                       &
  \colhead{LBLS$^{(b)}$}                     & \colhead{S$\&$S$^{(c)}$}          &
  \colhead{Other}                            & \colhead{$\theta_{\rm off}$}           &
  \colhead{$\tau_{K}$}                       & \colhead{SC$^{(e)}$} \\
  \colhead{number}                           & 
  \colhead{($^{h}$ $^{\rm mn}$ $^{\rm s}$)}  & \colhead{($^{\circ}$ $'$ $''$)}               & 
  \colhead{$(\%)$}                           & \colhead{$^{(\circ)}$}                    & 
  \colhead{$^{(\circ)}$}                         & \colhead{(Mag.)}                      &
  \colhead{}                                 & \colhead{}                            &       
  \colhead{ID$^{(d)}$}                     & \colhead{$^{(\circ)}$}                         &
  \colhead{}                                 & \colhead{$$}              \\
  \colhead{(1)}                              & 
  \colhead{(2)}                              & \colhead{(3)}                         & 
  \colhead{(4)}                              & \colhead{(5)}                         & 
  \colhead{(6)}                              & \colhead{(7)}                         &         
  \colhead{(8)}                              & \colhead{(9)}                         &   
  \colhead{(10)}                             & \colhead{(11)}                         &
  \colhead{(12)}                             & \colhead{(13)}     }
\startdata
0 &        5 35 16.2&  -5 21 54.2&  6.8 &  71.0& 1.3&0.48 &... &...   &...&...&0.27&0\\
1 &        5 35 15.9&  -5 21 51.2&  1.4 & 102.0& 6.1&0.47 &... &...   &...&75 &0.25&1\\
2 &        5 35 15.0&  -5 21 51.5&  2.6 &  81.0& 3.3&0.84 &... &...   &...&48 &0.81&1\\
3 &        5 35 14.7&  -5 21 53.6&  5.0 & 145.0& 1.7&1.69 &... &...   &...&71 &2.08&1\\
4 &        5 35 15.2&  -5 21 59.0&  5.1 & 110.0& 1.7&1.22 &... &...   &...&79 &1.38&1\\
5 &        5 35 15.0&  -5 22 02.8&  0.9 &  61.0& 9.5&0.19 &... &...   &...&...&-0.16&0\\
8 &        5 35 13.9&  -5 22 02.8&  7.8 & 144.0& 1.1&2.34 &... &...   &...&73 &3.06&1\\
9 &        5 35 13.2&  -5 21 57.1&  5.3 & 116.0& 1.6&1.09 &... &...   &...&77 &1.18&1\\
11&        5 35 14.6&  -5 22 09.0&  3.8 &  46.0& 2.3&0.34 &s   &...   &...&...&0.06&0\\
12&        5 35 13.9&  -5 22 09.6&  1.8 &  56.0& 4.8&0.53 &h   &... &P1089&19 &0.34&1\\
13&        5 35 16.1&  -5 22 12.0&  2.2 &  66.0& 3.9&1.29 &... &...   &...&46 &1.48&1\\
14&        5 35 16.0&  -5 22 13.9&  1.1 &  60.0& 7.8&0.80 &... &...   &...&37 &0.75&1\\
15&        5 35 15.3&  -5 22 16.9&  2.1 &  65.0& 4.1&1.21 &y   &...   &...&36 &1.36&1\\
16&        5 35 13.9&  -5 22 18.7& 11.2 & 108.0& 0.8&2.12 &... &...   &...&75 &2.73&1\\
17&        5 35 13.2&  -5 22 16.9&  3.0 &  42.0& 2.9&0.45 &a   &...   &...&...&0.22&0\\
18&        5 35 13.4&  -5 22 22.3&  2.6 &  24.0& 3.3&1.05 &b   &CB1   &...&9  &1.12&1\\
19&        5 35 13.7&  -5 22 20.9&  4.1 &  93.0& 2.1&0.61 &e   &...   &...&...&0.46&0\\
20&        5 35 13.9&  -5 22 22.8&  1.8 &  32.0& 4.8&0.74 &g   &...   &...&2  &0.66&1\\
21&        5 35 13.5&  -5 22 26.8&  1.3 &  62.0& 6.6&0.72 &c   &...   &...&30 &0.63&1\\
22&        5 35 16.2&  -5 22 22.0&  1.1 &  72.0& 7.8&0.94 &... &...   &...&52 &0.96&1\\
23&        5 35 15.8&  -5 22 21.7&  1.1 &  68.0& 7.8&1.10 &... &...   &...&48 &1.20&1\\
24&        5 35 15.3&  -5 22 25.5&  9.5 &  74.0& 0.9&0.83 &z   &...   &...&...&0.79&0\\
25&        5 35 15.2&  -5 22 24.4&  8.9 &  70.0& 1.0&0.46 &... &...   &...&...&0.24&0\\
26&        5 35 14.9&  -5 22 31.4& 15.4 &  51.0& 0.6&0.48 &u   &...   &...&...&0.27&0\\
27&        5 35 14.8&  -5 22 29.8& 20.7 & 141.0& 0.4&3.73 &... &CB4   &...&70 &5.14&1\\
30&        5 35 14.7&  -5 22 33.3&  1.3 &  58.0& 6.6&0.34 &t   &... &P1839&...&0.06&0\\
31&        5 35 14.4&  -5 22 32.5&  2.3 &  80.0& 3.7&2.37 &n   &n     &...&48 &3.10&1\\
33&        5 35 15.2&  -5 22 36.0&  9.2 & 141.0& 0.9&2.58 &x   &...   &...&68 & 3.42&1 \\
34&        5 35 14.9&  -5 22 38.2&  0.6 &  41.0&14.3&0.57 &v   &... &P1840&12 &0.40&1\\
35&        5 35 15.8&  -5 22 44.1&  1.8 &  28.0& 4.8&0.13 &... &...   &... &...& -0.25 &0 \\
36&        5 35 15.4&  -5 22 46.5&  2.2 &  26.0& 3.9&0.37 &... &...   &... &...&0.10  &0 \\
37&        5 35 14.7&  -5 22 47.4&  3.3 &  33.0& 2.6&1.37 &... &...   &... &4  & 1.60&1  \\
38 (BN)&   5 35 14.2&  -5 22 23.6& 18.0 & 115.0& 0.5&3.86 &BN  &...   &IRc1&82 &5.34 &1  \\
\enddata
\tablenotetext{(a)}{A precision $\sigma_{P}= \pm 0.3 \%$ can be assumed for these data.}
\tablenotetext{(b)}{\citet{lon82}}
\tablenotetext{(c)}{See \citet{sto98} and \citet{shu04}}
\tablenotetext{(d)}{ID:Identification Designation}
\tablenotetext{(e)}{SC:Selection Code, indicates the reliability of polarization data to probe magnetic fields.
SC $=$ 1 for reliable data. SC$=$ 0 otherwise.}
\end{deluxetable}

\clearpage

\begin{deluxetable}{cccccc}
\tablewidth{0pt}
\tabletypesize{\scriptsize}
\tablecaption{Means and dispersions of polarization and position angles at various
wavelengths and various scales in and around regions Orion A, OMC-1 and BN. \label{tabcomp}}
\tablehead{
 \colhead{}             &\colhead{Data set}  & \colhead{Large scale region}                 & 
 \colhead{OMC-1 region}      & \colhead{BN region}          \\

 \colhead{}                          & \colhead{}            & 
 \colhead{(Figure \ref{lsunp})}                   & 
 \colhead{(Figure \ref{compomc1})}   & \colhead{(Figure \ref{compbn})}  \\

 \colhead{}  & \colhead{}            & \colhead{($\approx 1.5^{\circ} \times 2.1^{\circ}$)}   & 
 \colhead{($\approx 21'\times17' $)}       & \colhead{($\approx 1.5' \times 1.5' $)}            \\

 \colhead{(1)}  & \colhead{(2)}            & 
 \colhead{(3)}  & \colhead{(4)}  & \colhead{(5)}           }
\startdata

$\overline{P}_{\rm V} \pm S_{P_{\rm V}}$ $(^{\circ})$$^{(a)}$                             &$^{(b)}$                         &1.83$\pm$2.06 ($N^{(e)}$=61)   &2.38$\pm$2.87  ($N=20$)            &$1.8\pm0.0$  ($N=1$) \\
$\overline{P}_{\rm V} \pm S_{P_{\rm V}}$ $(^{\circ})$$^{(a)}$                             &$(P \ge 1\%)$ $^{(c)}$           &2.87$\pm$2.27 ($N=34$)         &2.94$\pm$3.13  ($N=15$)            &... \\
$\overline{P}_{\rm V} \pm S_{P_{\rm V}}$ $(^{\circ})$$^{(a)}$                             &$(P<1\%)$ $^{(d)}$               &0.52$\pm$0.28 ($N=27$)         &0.68$\pm$0.22  ($N=5$)             &... \\
$\overline{P}_{\rm 2.2\mu m} \pm S_{P_{\rm 2.2\mu m}}$ $(^{\circ})$                       &$^{(b)}$                         &...                            &...                                &$4.89\pm5.47$  ($N=22$) \\  
$\overline{P}_{\rm 350 \mu m}\pm S_{P_{\rm 350 \mu m}}$ $(^{\circ})$                      &$^{(b)}$                         &...                            &2.65$\pm$1.48  ($N=470$)           &$2.18\pm0.42$  ($N=25$) \\
$\overline{\theta}_{\rm V} \pm S_{\theta_{\rm V}}$ $(^{\circ})$$^{(a)}$                   &$^{(b)}$                         &79.7$\pm$30.0 ($N=61$)         &63.6$\pm$38.9  ($N=20$)            &$107.5\pm0.0$  ($N=1$) \\
$\overline{\theta}_{\rm V} \pm S_{\theta_{\rm V}}$ $(^{\circ})$$^{(a)}$                   &($P \ge 1\%)$ $^{(c)}$           &89.6$\pm$27.8 ($N=34$)         &66.1$\pm$40.3  ($N=15$)            &... \\
$\overline{\theta}_{\rm V} \pm S_{\theta_{\rm V}}$ $(^{\circ})$$^{(a)}$                   &($P<1\%)$ $^{(d)}$               &67.3$\pm$28.3 ($N=27$)         &92.1$\pm$29.6  ($N=5$)             &... \\
$\overline{\theta}_{\rm 2.2\mu m} \pm S_{\theta_{\rm 2.2\mu m}}$ $(^{\circ})$             &$^{(b)}$                         &...                            &...                                &$84.6\pm$38.5  ($N=22$) \\        
$\overline{\theta}_{\rm 350 \mu m}\pm90^{\circ} \pm S_{\theta_{\rm 350 \mu m}}$$(^{\circ})$&$^{(b)}$                        &...                            &111.6$\pm$30.1  ($N=470$)          &$119.7\pm8.7$  ($N=25$) \\
\enddata
\tablenotetext{(a)}{Include data from Tables \ref{OMMDATA}, \ref{PVISDATA}, \ref{OTHERDATA} and \ref{OMC2DATA}.}
\tablenotetext{(b)}{No threshold on the degrees of polarization.}
\tablenotetext{(c)}{Only data with $P \ge 1\%$ are considered.}
\tablenotetext{(d)}{Only data with $P < 1\%$ are considered.}
\tablenotetext{(e)}{$N$: number of measurements in the set considered.}
\end{deluxetable}

\clearpage

\begin{deluxetable}{cccccc}
\tablewidth{0pt}
\tabletypesize{\scriptsize}
\tablecaption{Means and dispersions of polarization and position angles at various
wavelengths and various scales in and around regions OMC-2/3 and OMC-4. \label{tabcomp2}}
\tablehead{
 \colhead{}             &\colhead{Data set}  & \colhead{OMC-2/3 region}                 & 
 \colhead{OMC-2/3 inner region}      & \colhead{OMC-4 region}          \\

 \colhead{}                          & \colhead{}            & 
 \colhead{(Figure \ref{compomc23})}                   & 
 \colhead{(Figure \ref{compomc23zoom})}   & \colhead{(Figure \ref{compomc4})}  \\

 \colhead{}  & \colhead{}            & \colhead{($\approx 26' \times 30'$)}   & 
 \colhead{($\approx 3'\times5.5' $)}       & \colhead{($\approx 14' \times 14'$)}            \\

 \colhead{(1)}  & \colhead{(2)}            & 
 \colhead{(3)}  & \colhead{(4)}     & \colhead{(5)}       }
\startdata

$\overline{P}_{\rm V} \pm S_{P_{\rm V}}$ $(^{\circ})$$^{(a)}$                             &$^{(b)}$                         &2.15$\pm$2.50 ($N^{(e)}$=34)  &$1.00\pm1.03$  ($N=9$)    &1.20$\pm$1.27  ($N=13$) \\
$\overline{P}_{\rm V} \pm S_{P_{\rm V}}$ $(^{\circ})$$^{(a)}$                             &$(P \ge 1\%)$$^{(c)}$            &3.30$\pm$2.72 ($N=20$)        &$2.53\pm1.47$  ($N=2$)    &2.17$\pm$1.74  ($N=5$)  \\
$\overline{P}_{\rm V} \pm S_{P_{\rm V}}$ $(^{\circ})$$^{(a)}$                             &$(P<1\%)$ $^{(d)}$               &0.50$\pm$0.22 ($N=14$)        &$0.56\pm0.27$  ($N=7$)    &0.60$\pm$0.28  ($N=8$)  \\
$\overline{P}_{\rm 350 \mu m}\pm S_{P_{\rm 350 \mu m}}$ $(^{\circ})$                      &$^{(b)}$                         &1.79$\pm$1.16 ($N=137$)       &$1.41\pm1.26$  ($N=52$)   &1.66$\pm$0.39  ($N=19$) \\
$\overline{P}_{\rm 850 \mu m}\pm S_{P_{\rm 850 \mu m}}$ $(^{\circ})$                      &$^{(b)}$                         &2.82$\pm$1.62 ($N=251$)       &$1.98\pm1.00$  ($N=91$)   &...                     \\
$\overline{\theta}_{\rm V} \pm S_{\theta_{\rm V}}$ $(^{\circ})$$^{(a)}$                   &$^{(b)}$                         &66.3$\pm$34.6 ($N=34$)        &$46.9\pm29.7$  ($N=9$)    &54.5$\pm$36.3  ($N=13$) \\
$\overline{\theta}_{\rm V} \pm S_{\theta_{\rm V}}$ $(^{\circ})$$^{(a)}$                   &($P \ge 1\%)$ $^{(c)}$           &$71.1\pm37.7$ ($N=20$)        &$37.6\pm16.6$  ($N=2$)    &42.4$\pm$40.4  ($N=5$)  \\
$\overline{\theta}_{\rm V} \pm S_{\theta_{\rm V}}$ $(^{\circ})$$^{(a)}$                   &($P<1\%)$ $^{(d)}$               &$59.4\pm29.5$ ($N=14$)        &$49.6\pm33.1$  ($N=7$)    &39.5$\pm$33.9  ($N=8$)  \\
$\overline{\theta}_{\rm 350 \mu m}\pm90^{\circ} \pm S_{\theta_{\rm 350 \mu m}}$$(^{\circ})$&$^{(b)}$                        &44.7$\pm$24.6 ($N=137$)       &$48.7\pm32.7$  ($N=52$)   &75.7$\pm$12.6  ($N=19$) \\
$\overline{\theta}_{\rm 850 \mu m}\pm90^{\circ} \pm S_{\theta_{\rm 850 \mu m}}$$(^{\circ})$&$^{(b)}$                        &55.5$\pm$35.5 ($N=251$)       &$69.0\pm42.5$  ($N=91$)   &...                     \\
\enddata
\tablenotetext{(a)}{Include data from Tables \ref{OMMDATA}, \ref{PVISDATA}, \ref{OTHERDATA} and \ref{OMC2DATA}.}
\tablenotetext{(b)}{No threshold on the degrees of polarization.}
\tablenotetext{(c)}{Only data with $P \ge 1\%$ are considered.}
\tablenotetext{(d)}{Only data with $P < 1\%$ are considered.}
\tablenotetext{(e)}{$N$: number of measurements in the set considered.}
\end{deluxetable}

\end{document}